%% file: 2hdmreco.tex
\documentclass[11pt]{article}
\usepackage{epsf,amsmath,amssymb,graphicx,scalefnt,url,cite,color}
\title{
\vspace{-4em}
\begin{flushright}
{\small 
LHCHXSWG-2013-001\vspace{0.5em}\\
KA--TP--41--2013\\
LU TP 13--44 \\
PSI--PR--13--17\\
WUB/13-19\\
}
\vspace{2em}
\end{flushright}
\Large LHC Higgs Cross Section Working Group\\
\vspace{0.5em}
Interim recommendations for the evaluation of Higgs production cross sections and
  branching ratios at the LHC in the Two-Higgs-Doublet~Model}
\author{%
R.\,Harlander$^1$,
M.\,M\"uhlleitner$^2$,
J.\,Rathsman$^3$,
M.\,Spira$^4$,
O.\,St\r{a}l$^5$\\[2em]
$^1$ {\it Fachbereich C, Bergische Universit\"at Wuppertal,}\\ 
{\it 42097 Wuppertal, Germany}\\[0.5em]
$^2$ {\it Institut f\"ur Theoretische Physik,
Karlsruher Institut f\"ur Technologie KIT,}\\{\it 76131 Karlsruhe, Germany}\\[0.5em]
$^3$ {\it Department of Astronomy and Theoretical Physics}\\ {\it Lund University, SE-223 62, Lund, Sweden}\\[0.5em]
$^4$ {\it Paul Scherrer Institut, CH--5232 Villigen PSI, Switzerland}\\[0.5em]
$^5$ {\it The Oskar Klein Centre,
Department of Physics}\\{\it Stockholm University, SE-106 91 Stockholm, Sweden}\\
}

\begin{document}

\maketitle

\begin{abstract}
\noindent
In this note we give interim recommendations on how to evaluate LHC cross
sections for (neutral) Higgs production and Higgs branching ratios in
the general (CP-conserving) Two-Higgs-Doublet Model (2HDM). The current
status of available higher-order corrections to Higgs production and
decay in this model is discussed, and the existing public codes
implementing these calculations are described. Numerical results are
presented for a set of reference scenarios, demonstrating the very good
agreement between the results obtained using different programs.
\end{abstract}
\newpage


\section{Introduction}
The Two-Higgs-Doublet Model (2HDM) is one of the simplest extensions of
the Standard Model (SM) Higgs sector. It can be useful both as a theoretical tool to explore
the phenomenology of an extended Higgs sector, and to interpret experimental results from searches for additional Higgs bosons. A comprehensive review of this model is given in Ref.~\cite{Branco:2011iw}.
The existence of a Higgs
boson with mass around $M_H=125$\,GeV and couplings compatible with the
SM predictions already leads to severe restrictions of the 2HDM parameter space.

Precise analyses of the experimental data on the basis of the 2HDM
require solid predictions of the relevant observables in this model. Two
of the most important ones are the production cross sections and the
branching ratios of the Higgs bosons. In this note, we provide a brief
overview of the theory status of these quantities. We describe the
available radiative corrections, and the most convenient tools that
allow one to obtain numerical predictions for specific sets of 2HDM
parameters. We also comment on current theoretical short-comings and
open issues. The results are summarized in the form of interim recommendations for the evaluation of Higgs production cross sections and branching ratios in the 2HDM that represent---to the best of our knowledge---the current best practice based on available tools.


\section{Theory recommendations}


\subsection{Cross sections}\label{sec:xsec}


\subsubsection{Higgs production in the Standard Model}\label{sec:xsecsm}

The dominant production cross section for Higgs bosons in the Standard
Model is gluon fusion. The coupling of the gluons to the Higgs boson is
mediated predominantly by a top quark loop, while loops from other
quarks are suppressed by their Yukawa couplings. The effect of bottom
quark loops amounts to about $-6\%$, charm quark loops reach about
$-1\%$.  QCD corrections to the gluon fusion cross section are large,
reaching more than a factor of two at NNLO
QCD\,\cite{Harlander:2002wh,Anastasiou:2002yz,Ravindran:2003um}. While
the NLO QCD corrections can be calculated for general quark
mass\,\cite{Graudenz:1992pv,Spira:1993bb,Spira:1995rr}, the NNLO terms
are available only in the heavy-top approximation. The validity of this
approximation has been checked to be better than 1\% for Higgs masses
below 300\,GeV\,\cite{Marzani:2008ih,
Harlander:2009my,Harlander:2009mq,Pak:2009dg,Pak:2011hs}.

Various effects beyond fixed order perturbative corrections have been
studied, but they are not included in {\tt SusHi}
\cite{Harlander:2012pb} or {\tt HIGLU} \cite{higlu} (see below). They
include soft gluon resummation\,\cite{Catani:2003zt,Ravindran} (which,
as has been argued, can be accounted for by choosing the
renormalization/factorization scale to
$\mu=M_H/2$\,\cite{Anastasiou:2008tj}), and partial or approximate
N$^3$LO results\,\cite{Ravindran,Moch:2005ky,Ball:2013bra}.

Electroweak corrections are also
known\,\cite{Djouadi:1994ge,Actis:2008ug,Degrassi:2004mx,Aglietti:2004nj},
and a calculation of the mixed electroweak/QCD
effects\,\cite{Anastasiou:2008tj} -- albeit in the unphysical limit
$M_W>M_H$ -- suggests that they approximately factorize from the QCD
effects, since the latter are dominated by soft-gluon radiation.


\subsubsection{Higgs production in the 2HDM}\label{sec:xsec2hdm}

The calculation of the gluon fusion cross section for scalar Higgs
bosons in the 2HDM differs from the SM only by the Higgs couplings.
Therefore the QCD corrections in the 2HDM can be obtained quite
trivially by separately rescaling the SM expressions for the top- and
bottom-loop induced amplitudes. Note, however, that the bottom-loop
contribution can be enhanced in the 2HDM.  Since NNLO corrections are
only known for the top-loop contribution, the accuracy of the prediction
decreases with increasing values of the ratio $g_b/g_t$, where $g_q$ is
the $Hq\bar q$ coupling. An analogous procedure can be applied for the
pseudoscalar Higgs boson if the individual scalar amplitudes are
replaced by the corresponding pseudoscalar ones.

Note also that electroweak effects are only fully available in the SM.
However, the electroweak corrections involving light quarks (i.e.\,all
but the top quark) are known separately\,\cite{Aglietti:2004nj}; since
they are proportional to the $VVH$ coupling, where $V\in\{W,Z\}$, they
can also be rescaled quite easily from the SM to the 2HDM. They do not
contribute to pseudoscalar Higgs boson production, since the
pseudoscalar does not couple to $W,Z$ bosons at tree level.

If the ratio $g_b/g_t$ is sizable, another production mechanism comes
into play, namely associated $\phi b\bar b$ ($\phi=h,H,A$)
production\,\cite{Dicus:1988cx,Campbell:2002zm,Dicus:1998hs,
  Maltoni:2003pn,Dittmaier:2003ej,Dawson:2003kb}. Its fully inclusive
cross section can be calculated through NNLO in the 5-flavor scheme
(5FS)\,\cite{Harlander:2003ai}, which effectively corresponds to the
process $b\bar b\to \phi$. The corresponding 4-flavor-scheme (4FS)
numbers can be generated by using the grids for the corresponding MSSM
cross sections \cite{4fsgrids} and inserting the appropriate factors for
the bottom Yukawa couplings. Both calculations can finally be merged by
using the Santander matching \cite{santander}.


\subsubsection{The invalidity of global K-factors}

As described in Section\,\ref{sec:xsec2hdm}, the result for the gluon
fusion cross section $gg\to \phi$, where $\phi$ represents both the
light $h$ and the heavy $H$ CP-even Higgs boson of the 2HDM, can be
obtained by replacing the Yukawa couplings in the SM amplitude by their
2HDM values. One may wonder whether it is sufficient for the inclusive
Higgs boson production cross section to apply this rescaling only at LO
and take the QCD corrections into account as an overall
factor:
\begin{equation}
\sigma^{\rm{approx}}_{\rm{2HDM}} 
\stackrel{?}{=}
 \frac{\sigma^{\rm{NNLO}}_{\rm{SM}}}{\sigma^{\rm{LO}}_{\rm{SM}}} \left\{
g_t^2 \sigma^{\rm{LO}}_{tt}
+ g_t g_b \sigma^{\rm{LO}}_{tb}
+ g_b^2 \sigma^{\rm{LO}}_{bb}\right\}\,,
\label{eq:approx}
\end{equation}
where
$\sigma^{\rm{LO}}_{\rm{SM}}=\sigma^{\rm{LO}}_{tt}+\sigma^{\rm{LO}}_{tb}+\sigma^{\rm{LO}}_{bb}$,
with $\sigma_{tt}$ and $\sigma_{bb}$ the contributions due to the top
and the bottom loop, respectively, and $\sigma_{tb}$ arising from the
top-bottom interference. The coefficients $g_{t,b}$ denote the top and
bottom Yukawa couplings, normalized to the SM ones. They depend only on
the Higgs mixing angles $\alpha$ and $\beta$ in general. This approach
assumes that the relative QCD corrections are about the same for all
three parts. This hypothesis, however, can be checked explicitly by
comparing the individual numbers at LO, NLO and NNLO. In
Table\,\ref{tb:ggf8} the results (obtained with {\tt
  HIGLU}\,\cite{higlu} and cross-checked with {\tt
  SusHi}\,\cite{Harlander:2012pb}) are displayed for the LHC at 8 TeV
c.m.\ energy and in Table\,\ref{tb:ggf14} for 14 TeV.  While the pure
top-loop contributions exhibit the known large K-factors, this is not
the case for the $tb$-interference nor for the pure bottom loops. The
approximation of Eq.~(\ref{eq:approx}) is therefore off by up to 50\%
for cases in which the bottom loops provide the dominant contributions.
Note in particular that the QCD corrections to the interference term
$\sigma_{tb}$ are always moderate and can be of either sign.  Assuming a
``global K-factor'' is therefore clearly invalid; instead, the full NLO
(NNLO) calculation has to be used in order to obtain reliable results
beyond LO. Fortunately, they can easily be obtained with the help of
{\tt SusHi} \cite{Harlander:2012pb} or {\tt HIGLU} \cite{higlu}.

\begin{table}[hbtp]
\renewcommand{\arraystretch}{1.2}
\begin{center}
\input{tables/table-global-k-8}
\renewcommand{\arraystretch}{1.0}
\caption[]{\label{tb:ggf8} \it SM Higgs production cross sections via gluon
fusion at leading order (LO), next-to-leading order (NLO) and including
the next-to-next-to-leading order (NNLO) corrections to the top loops in
the heavy-top limit on top of the next-to-leading order predictions
using MSTW2008 PDFs for $\sqrt{s}=8$ TeV. The renormalization and
factorization scales have been chosen as $\mu_R=\mu_F=M_H/2$.}
\end{center}
\end{table}

\begin{table}[hbtp]
\renewcommand{\arraystretch}{1.2}
\begin{center}
\input{tables/table-global-k-14}
\renewcommand{\arraystretch}{1.0}
\caption[]{\label{tb:ggf14} \it Same as Table\,\ref{tb:ggf8}, but for $\sqrt{s}=14$ TeV.}
\end{center}
\end{table}



\subsection{Transverse momentum distribution}

\begin{figure}[hbt]
\centering
\includegraphics[width=0.9\textwidth,viewport=100 480 520 800]{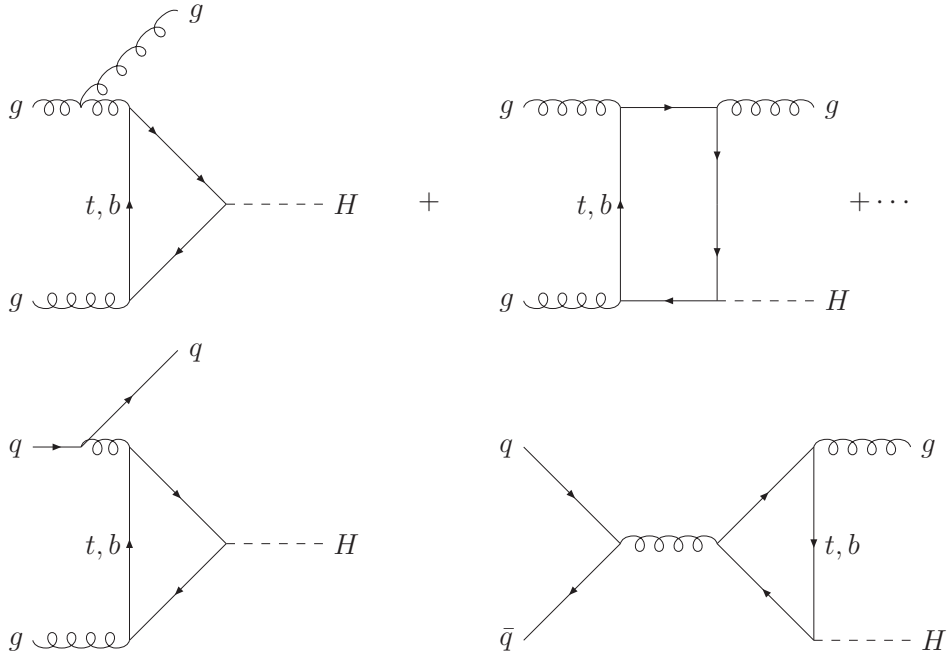}
\caption[]{\it \label{fg:hptdia} Typical Feynman diagrams contributing to
Higgs + jet production emerging from $gg,qg$ and $q\bar q$ initial
states.}
\end{figure}

Just like the inclusive cross section, the transverse-momentum ($p_T$)
distribution of the SM Higgs boson in gluon fusion is predominantly
mediated by top-quark loops, giving rise to the process $gg\to gH$,
supplemented by the subleading channels $gq\to qH$ and $q\bar q\to gH$
\cite{ptfix}, see Fig.~\ref{fg:hptdia}.  However, since the bottom quark
contribution amounts to about $-6\%$ in the inclusive cross section, we
also expect a similar effect on the $p_T$-distribution. At leading
order, the bottom quark contribution is only sizable for small $p_T$,
while for larger $p_T$ values it can safely be neglected. The NLO
corrections to the $p_T$-distribution are only known in the heavy-top
limit\,\cite{ptnlo}\footnote{In this approximation, the $gg$-component
  for this process was even calculated at NNLO\,\cite{ptnnlo}.},
including subleading terms in the inverse top mass at
NLO\,\cite{ptnlomt}. Similar to the inclusive cross section, the QCD
corrections are large, nearly doubling the rate at finite $p_T$ values.

It is well-known that the pure LO and NLO results diverge towards
$p_T\to 0$. The small $p_T$ region thus requires a soft gluon
resummation in order to achieve a reliable prediction. This resummation
has been performed systematically for the top quark loops in
Ref.\,\cite{ptresum}, neglecting finite top mass effects at NLO. Since
soft gluon effects factorize, the top mass effects at small $p_T$ are
well approximated by the LO mass dependence for small Higgs
masses\,\cite{Alwall:2011cy,Bagnaschi:2011tu,Mantler:2012bj}. As long as
the top-loop contribution dominates the cross section, the only limiting
factor of the NLO+NNLL result as implemented in {\tt
  HqT} or {\tt HRes}\,\cite{grazzinicodes} is thus the
heavy-top approximation of the NLO corrections which affects the small
$p_T$ range for large Higgs masses and the large $p_T$ region.

Recently, bottom quark contributions have been included in the
predictions for the resummed
$p_T$-distributions\,\cite{Bagnaschi:2011tu,Alwall:2011cy,%
  Grazzini:2013mca,Mantler:2012bj,Banfi:2013eda}. It was
shown\,\cite{Grazzini:2013mca,Banfi:2013eda} that factorization of the
soft gluon effects fails in this case if the resummation scale is chosen
to be of the order of the Higgs mass, as it is done for the top-loop
contributions.

In Ref.\,\cite{Grazzini:2013mca}, it was therefore suggested to
introduce an additional resummation scale of the order of $m_b$ for the
terms involving bottom quark loops.  Due to the relatively large
separation of scales ($m_b$, $M_H$, $p_T$), certain potentially large
logarithms remain, which are still treated perturbatively in this
approach. The relative theoretical uncertainty is thus much larger than
for the top quark terms.  This adds to the fact that the bottom loop
contributions can only be included at LO+LL, since the QCD corrections
to the $p_T$-distribution are unknown. Due to the small bottom Yukawa
coupling, the impact on the overall theoretical uncertainty is rather
modest in the SM though.

This is in contrast to 2HDM scenarios with a largely enhanced bottom
Yukawa coupling. The resulting theoretical uncertainty of the
differential cross section $d\sigma/dp_T$ could be of the order of
$100\%$ in the whole $p_T$ region. Analogously large uncertainties also
arise for the shape of the $p_T$ distribution, i.e.~the normalized
differential cross section $1/\sigma~d\sigma/dp_T$, since unresummed
large logarithms at small $p_T$ typically have an impact on the
high-$p_T$ region due to the unitarity constraint given by the total
cross section. For the time being, we recommend the use of the
implementation of mass effects in the {\tt POWHEG box}
\cite{Bagnaschi:2011tu} and {\tt MC@NLO} \cite{Frixione:2002ik,mcatnlo}
for the Higgs $p_T$ distributions, since both codes include LO quark
mass effects consistently.

Let us now turn to the transverse momentum distribution in associated
production of a Higgs boson with bottom quarks.  Similar to the
inclusive cross section, the $p_T$ distribution of the Higgs bosons can
also be obtained both in the 4FS and the 5FS~\cite{bbhlh}. In the 5FS,
the NLO prediction (i.e., without resummation) has been calculated in
Ref.\,\cite{Harlander:2010cz}. Note that in the 5FS the Higgs can be
produced at finite $p_T$ only when a jet is emitted from the initial
state, similar to the situation in gluon fusion. Imposing the
restriction that this jet be due to a $b$-quark, the NLO result was
found in Ref.\,\cite{Campbell:2002zm,Harlander:2010cz} and is
implemented in {\tt MCFM}\,\cite{mcfm}. Also the case with two final
state $b$ quarks was considered there, albeit only at LO in the 5FS. The
effect of a $p_T$-, jet-, or $b$-jet-veto through NNLO was considered in
Ref.\,\cite{Harlander:2011fx}; a fully differential Monte Carlo program
for $b\bar bH$ production in the 5FS was presented in
Ref.\,\cite{Buehler:2012cu}.  None of these results are applicable for
small $p_T$ where logarithms spoil the perturbative convergence. The
LO+LL resummation of these terms, on the other hand, can be found in
Ref.\,\cite{Belyaev:2005bs}.

In the 4FS calculation, the Higgs boson can be produced at finite $p_T$
already at LO. One could attempt to match the NLO
4FS\,\cite{Dittmaier:2003ej,Dawson:2003kb} with the 5FS
results\,\cite{Harlander:2010cz,Campbell:2002zm} along the lines of
Ref.\,\cite{santander} in order to obtain the most accurate prediction
of the Higgs $p_T$ distribution. This holds independent of the number of
$b$-tags in the final state: for zero and one $b$-tag, both 4FS and 5FS
results are formally of NLO, for two $b$-tags, the 5FS result is
available only at LO, while the 4FS result is still valid through NLO.
Note, however, that even in the latter case, both approaches include
partly complementary terms because the arrangement of the perturbative
series is different in the 4FS and the 5FS.

Unfortunately, we are not aware of any attempts to match 4FS and 5FS
results for the Higgs $p_T$ distribution in associated $b\bar bH$
production. Even public tools for the calculation of the general NLO or
resummed Higgs $p_T$ distribution in either of the two schemes are
unavailable to our knowledge. Given sufficient demand from the
experimental community, we would expect this to change quite quickly
though. Until then, we recommend to contact the authors of the papers
referred to in this section in order to obtain predictions for
particular sets of parameters, or to resort to tools like {\tt
  MCFM}\,\cite{mcfm} or {\tt aMC@NLO}\,\cite{amcnlo}, for specific
purposes.


\subsection{Branching ratios}

The decay widths and branching ratios of the 2HDM Higgs bosons are
usually calculated at leading order in the 2HDM parameters. This is the
case for both public codes, {\tt 2HDMC}~\cite{Eriksson:2009ws} and {\tt
HDECAY} \cite{Djouadi:1997yw,fortsch}, which are available to calculate
the 2HDM decays.  Using tree-level relations to convert between
different parametrizations (e.g.~physical masses and quartic couplings),
the results obtained with input in different bases are formally
equivalent.  However, unlike the case of the SM with $M_H=125$~GeV or
the MSSM, there is no automatic protection in the 2HDM against
arbitrarily large quartic couplings, which may lead to a violation of
perturbativity of the couplings and tree-level unitarity. This should be
kept in mind when calculating decay widths involving triple-Higgs
couplings, such as $h\to \gamma\gamma$ or $H\to hh$. To ensure
reasonable theoretical properties of the model, we recommend that
positivity of the Higgs potential, perturbativity of the couplings and
tree-level unitarity is applied as explicit constraints on the 2HDM
scenarios in all analyses.

Higher-order SM electroweak corrections do not factorize from the LO
result and cannot be readily included for the 2HDM. There is currently
no calculation of the electroweak corrections to the 2HDM Higgs decays
available in usable format. We therefore recommend that electroweak
corrections are not included in the calculation of 2HDM branching
ratios, and in particular that no corrections obtained for the SM are
applied. This introduces unavoidable uncertainties, which can be
estimated from the size of the known electroweak corrections within the
SM to be up to 5--10\% for several partial decay widths. Differences of this magnitude
 compared to the most precise values for the SM are therefore expected even in the decoupling limit.
 
A consistent comparison of 2HDM predictions in the decoupling limit to SM values furthermore requires that the limit is taken properly so that no residual 2HDM effects are present, e.g.~from $H^\pm$ loop contributions to $h\to \gamma\gamma/Z\gamma$. Using the physical Higgs mass basis as an example (see below), SM-like decays for the lightest 2HDM Higgs boson can be achieved by choosing $M_h\sim 125$~GeV, $\sin(\beta-\alpha)=1$, $M_H, M_A, M_{H^\pm} \gg v$, and the value of $M_{12}^2$ which makes the $hH^+H^-$ coupling vanish.

Unlike the electroweak corrections, many of the QCD corrections (which
typically are numerically significant) \emph{do} factorize, and can
therefore be taken over directly from the corresponding SM or MSSM
calculations. For numerically relevant branching ratios, we recommend
that QCD corrections are taken into account to the highest degree of
accuracy that is practically possible.

Schematically, the widths for the SM-type decays to quark pairs and
vector bosons are obtained at leading order from their SM equivalents by rescaling the
interaction vertices with the corresponding 2HDM factors. The
loop-mediated decay to gluons also proceeds as in the SM, with the
appropriate rescaling of the Yukawa couplings. For the remaining decays
it is necessary in addition to take 2HDM-specific contributions into account. Even
with the restriction to models which respect perturbativity of the
couplings and tree-level unitarity, it is not possible to {\`a}-priori
assume that any of these contributions are small over the 2HDM parameter
space. This is true in particular for the decays to vector bosons and
Higgs bosons, $\phi_i \to \phi_j V$ $(V=W/Z)$, or Higgs-to-Higgs-Higgs
decays, $\phi_iÊ\to \phi_j\phi_k$, which can both be important when
kinematically allowed. The resulting decay patterns could therefore
differ substantially from that of a SM-like Higgs boson with the same
mass. 

We recommend that a full 2HDM calculation of the branching ratios
is always performed for each parameter point under study, as can
easily be obtained using either {\tt 2HDMC} \cite{Eriksson:2009ws} or {\tt
HDECAY} \cite{Djouadi:1997yw,fortsch}. This is important to get both the
individual contributions (including higher-order corrections) and the total width (and, therefore, the
branching ratios) correct. It is \emph{not} recommended to rely on any
simplified approximations in this respect.



\section{Available programs}


\subsection{Comparison of {\tt HIGLU} and {\tt SusHi}}

The order of the QCD corrections to the gluon fusion cross section
implemented in {\tt SusHi}\,\cite{Harlander:2012pb} and {\tt
  HIGLU}\,\cite{higlu} is identical, although they are evaluated in
completely independent and partly quite different ways.  They include
the full NLO QCD corrections, taking into account all quark mass
effects, for the top-, bottom- and charm-loop contributions. NNLO QCD
corrections are included only for the top quark by applying the
heavy-top limit.  Both {\tt SusHi} and {\tt HIGLU} allow one to modify
the top and bottom Yukawa couplings through the input file.  For the SM,
both programs also apply the electroweak correction factor, assuming
full factorization of the QCD effects.

{\tt SusHi} also evaluates the cross section for $b\bar b\to \phi$
($\phi\in\{h,H,A\}$) through NNLO QCD within the 5-flavor scheme. It
also provides a link to {\tt 2HDMC} which allows for a number of
different parametrizations of the 2HDM, immediately passes on the
correct values of the Higgs couplings to {\tt SusHi}, and evaluates the
Higgs decay rates. {\tt SusHi} takes into account electroweak
corrections to the 2HDM cross section as induced by light-fermion loops,
see Section\,\ref{sec:xsec2hdm}.

{\tt HIGLU} is linked to {\tt HDECAY} thus allowing for a consistent use
together with the 2HDM extension of {\tt HDECAY}.


\subsection{Comparison of {\tt 2HDMC} and {\tt HDECAY}}


\underline{\bf The C++ code {\tt 2HDMC}} \\ 
{\tt 2HDMC}
\cite{Eriksson:2009ws} is a public C++ code available for download from
     {\sc HepForge} through \url{http://2hdmc.hepforge.org/}. It has been
     developed from the start as a tool for studies in the
     (CP-conserving) two-Higgs-doublet model. As such, it contains an
     implementation of the general form for the 2HDM potential
     \cite{Davidson:2005cw}. There are a number of different options
     available to specify the model parameters, which include:\newline
     \underline{\emph{Generic} basis}
\begin{eqnarray}
\begin{array}{rl}
\mbox{Quartic potential couplings:} & \lambda_1, \lambda_2,\lambda_3,\lambda_4,\lambda_5,\lambda_6,\lambda_7 \\
\mbox{Soft $Z_2$-breaking mass:} & M_{12}^2\; (\mathrm{GeV}^2) \\
\mbox{Ratio of vacuum expectation values (basis):} & \tan\beta \\
\end{array}
\nonumber          
\end{eqnarray} 
\underline{\emph{Physical} basis}
\begin{eqnarray}
\begin{array}{rl}
\mbox{Higgs boson masses:} & M_h, M_H, M_A,  M_{H^\pm}\; (\mathrm{GeV}) \\
\mbox{Basis invariant scalar mixing:} & \sin(\beta-\alpha) \\
\mbox{$Z_2$-breaking quartic couplings:} & \lambda_6, \lambda_7 \\
\mbox{Soft $Z_2$-breaking mass:} & M_{12}^2\; (\mathrm{GeV}^2) \\
\mbox{Ratio of vacuum expectation values (basis):} & \tan\beta \\
\end{array}
\nonumber          
\end{eqnarray} 
For details on the definitions of these input parameters, see \cite{Eriksson:2009ws}. Note that {\tt 2HDMC} uses the convention $-\frac{\pi}{2} \leq \beta-\alpha \leq \frac{\pi}{2}$, and that the \emph{physical} basis (with $\lambda_6=\lambda_7=0$) is equivalent to the input used by {\tt HDECAY} (see below). A model specified using one input can be translated between different parameter bases by {\tt 2HDMC}. 

Independent of the choice of model parametrization, {\tt 2HDMC} allows
the 2HDM Yukawa sector to be specified in the most general way as full
($3\times 3$) Yukawa matrices using the language of
\cite{Davidson:2005cw}. The case of (softly-broken) $Z_2$-symmetric 2HDM
\emph{types} I--IV, as defined in \cite{Barger:1989fj}, is also
available. These recommendations concern only this class of models.

{\tt 2HDMC} contains functions to test for consistency of the specified
model parameters, such as vacuum stability (at the input scale),
tree-level unitarity, and perturbativity of the couplings. It can also
evaluate one-loop contributions to the oblique electroweak parameters
and direct experimental constraints through an interface to the code
{\tt HiggsBounds} \cite{Bechtle:2013wla}. The Higgs decay modes are
calculated in {\tt 2HDMC} using the full 2HDM parametrization and
a general Yukawa sector, including tree-level FCNCs in models where they
are present. 

\underline{\bf The Fortran code {\tt HDECAY}} \\
The Fortran code {\tt HDECAY} \cite{Djouadi:1997yw,fortsch} has been
extended to include the computation of the Higgs boson decay widths in
the framework of the 2HDM. The input parameters to be specified in the
input file {\tt hdecay.in} are
\begin{eqnarray}
\begin{array}{ll}
\mbox{the mixing angle:} & \alpha \\
\mbox{the ratio of the vacuum expectation values:} & \tan\beta \\
\mbox{the mass values of the five Higgs bosons:} & M_h, M_H, M_A,
M_{H^\pm}  \; (\mathrm{GeV})\\
\mbox{the mass parameter squared:} & M_{12}^2 \; (\mathrm{GeV}^2)
\end{array}
\nonumber          
\end{eqnarray} 
and a flag specifying the type of the 2HDM. 
The models type I--IV are implemented. From the input parameters {\tt
HDECAY} calculates the couplings which are needed in the computation of
the decay widths. With the appropriate coupling replacements according
to the various 2HDM's, the decay widths are the same as for the MSSM
Higgs boson decays, which are already included in the program. Only the
SUSY particle contributions in the loop-mediated decays and the decays
into SUSY particles as well as higher order corrections due to SUSY
particle loops have been turned off for the 2HDM case.

\underline{\bf Decay modes} \\
The implemented decay widths and higher order corrections are specified
in the following for both codes, {\tt 2HDMC} (version 1.6.2) and {\tt HDECAY} (version 6.00). 

\noindent
\underline{Decays into quark pairs:}
The QCD corrections factorize with respect to the tree-level amplitude
and can therefore be taken over from the SM case. For
the neutral Higgs decays the fully massive NLO corrections near
threshold \cite{nearthreshold} and the massless ${\cal O} (\alpha_s^4)$
corrections far above threshold \cite{abovethreshold,chetyrkin,baikov}
have been included in {\tt HDECAY}. They are calculated in terms of the running quark
masses and of the running strong coupling constant in order to resum
large logarithms. The QCD corrections to the charged Higgs boson decays
have been taken from \cite{Djouadi:1994gf}. Note that for the higher
order corrections to the decay widths only QCD corrections are taken
into account, as the electroweak corrections cannot be adapted from the
SM case. For the decays of the heavier neutral Higgs bosons into a top
quark pair, in {\tt HDECAY} off-shell decays below threshold have been
implemented as well as for the decays of a charged Higgs boson into a
top-bottom quark pair \cite{offshellphi}. 

In {\tt 2HDMC} all decays into on-shell light quark pairs are included.
Far above the threshold, the dominant QCD corrections for the decays to
pairs of strange, charm, and bottom quarks are taken into account by
absorbing large logarithms into the running quark masses appearing in
the Yukawa couplings at a scale set by the decaying Higgs boson mass
\cite{nearthreshold}. QCD corrections in this limit are included to
NNLO, $\mathcal{O}(\alpha_s^2)$ \cite{abovethreshold}, for the neutral Higgs decays.
For the decay to
$t\bar{t}$, in {\tt 2HDMC} threshold corrections of
$\mathcal{O}(\alpha_s)$ are calculated according to
\cite{nearthreshold}, keeping the full dependence on $m_t$. Below
threshold, the contributions from the off-shell decay $h/H/A\to
t\bar{t}^{(*)} \to t\bar{b}W^- + \mathrm{c.c.}$ are included according to
\cite{offshellphi}. The calculations of charged Higgs decays in {\tt 2HDMC} contain the $\mathcal{O}(\alpha_s)$ QCD corrections from \cite{Djouadi:1994gf}, with full mass dependence near threshold for $H^+\to tb$. Below threshold, the three-body decay $H^+\to t^{(*)}\bar{b}\to \bar{b}bW^+ + \mathrm{c.c.}$ is included according to \cite{offshellphi}.

\noindent
\underline{Decays into gluons:}
The QCD corrections to the neutral Higgs boson decays into gluons, a
loop-induced process already at leading order, can be taken over from
the SM, respectively, the MSSM.  They have been included up to N$^3$LO
in {\tt HDECAY}.  While for the SM at NLO the full quark mass dependence
\cite{Spira:1995rr} is available, for the 2HDM the corrections have been
taken into account in the limit of heavy-quark loop particle masses
\cite{nloggqcd,Chetyrkin:1997un,Baikov:2006ch}.

As in {\tt HDECAY} the contributions from all colored fermions are
included to one loop in {\tt 2HDMC}. For the contribution from the top
quark loop, $\mathcal{O}(\alpha_s^2)$ NNLO QCD corrections are included
in the heavy-top limit \cite{Chetyrkin:1997un}.

\noindent
\underline{Decays into photon pairs:} The decay to a photon pair is
loop-mediated, with the two most important SM contributions being due to
the top quark and $W$ boson loops. In the 2HDM, there is also a $H^+$
contribution which becomes numerically significant in some cases. The
bottom loop becomes relevant for the photonic decay rate in scenarios
with enhanced bottom Yukawa couplings.  For this decay, a special
running quark mass is employed to capture QCD corrections
\cite{Djouadi:1993ji} in {\tt 2HDMC} (the same as used in {\tt HDECAY}).
Explicit corrections to the top quark contribution in the heavy-top
limit (neglecting the mass dependence) are included to
$\mathcal{O}(\alpha_s)$ \cite{hgagalim}. In the pseudoscalar case only
heavy charged fermion loops contribute.

In {\tt HDECAY} the neutral Higgs boson decays into a photon pair have
been implemented at NLO QCD including the full mass dependence for the
quarks \cite{Spira:1995rr,Djouadi:1993ji,hgagaqcd}.

\noindent
\underline{Decays into $Z\gamma$:}
The loop induced decays of scalar Higgs bosons into $Z\gamma$ are
mediated by $W$ and heavy charged fermion loops, while the pseudoscalar
decays proceed only through charged fermion loops. The QCD corrections
to the quark loops are numerically small \cite{nloZga} and have not been
taken into account in {\tt 2HDMC} nor in {\tt HDECAY}.

\noindent
\underline{Decays into massive gauge bosons:}
The decay widths of the scalar Higgs bosons into massive gauge bosons
$\phi \to V^{(*)}V^{(*)}$ ($V=W,Z$) are the same as the SM decay width
after replacing the SM Higgs coupling to gauge bosons with the
corresponding 2HDM Higgs coupling. The option of double off-shell decays
\cite{Cahn:1988ru} has been included in {\tt 2HDMC} and {\tt HDECAY}.
The pseudoscalar Higgs boson does not decay into massive gauge bosons at
tree level. 

\noindent
\underline{Decays into Higgs boson pairs:}
The heavier Higgs particles can decay into a pair of lighter Higgs
bosons. This is a feature of the 2HDM which does not exist in the SM.
Due to more freedom in the mass hierarchies compared
to the MSSM case, the following Higgs-to-Higgs decays are
possible and have been included in {\tt 2HDMC} and {\tt HDECAY},
\begin{eqnarray}
h \to AA^{(*)} \,, \qquad H \to hh^{(*)} \,, \qquad H \to AA^{(*)}  \;.
\end{eqnarray}
In addition the decays into a charged Higgs boson pair are possible,
\begin{eqnarray}
h \to H^+ H^- \,, \quad \quad H \to H^+ H^- \;.
\end{eqnarray}
In both programs all decays are calculated at leading order, retaining
the full parameter dependence of the 2HDM for the trilinear couplings.
The contributions from final states with an off-shell scalar or
pseudoscalar, which can be significant, have been taken into account in
{\tt HDECAY} \cite{offshellphi}.  It is important to note that the
partial width for these decays can grow very large for parameter points
that do not respect the requirements of perturbativity and tree-level
unitarity.

\noindent
\underline{Decays into a gauge and Higgs boson pair:}
The Higgs boson decays into a gauge and a Higgs boson \cite{fortsch},
which have been implemented in both programs, including the possibility
of off-shell gauge bosons \cite{offshellphi}, are given by
\begin{eqnarray}
\begin{array}{lllllllll}
h &\to& A Z^{(*)} \,, &\quad h &\to& H^\pm W^{\mp (*)}\,, \\
H &\to& A Z^{(*)} \,, &\quad H &\to& H^\pm W^{\mp (*)}\,, \\
A &\to& h Z^{(*)} \,, &\quad A &\to& H Z^{(*)} \,, &\quad A &\to&
H^\pm W^{\mp (*)} \,,
\\  
H^\pm &\to& h W^{\pm (*)}\,, &\quad H^\pm &\to& H W^{\pm (*)}\,, &\quad H^\pm
&\to& A W^{\pm (*)} \;.
\end{array}
\end{eqnarray}
They have been implemented at leading order and include the
contributions of off-shell $W$ and $Z$ bosons below threshold
\cite{offshellphi}.



\section{Numerical results}


The programs dicussed in these recommendations, {\tt SusHi}, {\tt
  HIGLU}, {\tt 2HDMC} and {\tt HDECAY}, all implement the most general
(CP-conserving) version of the 2HDM with a softly-broken $Z_2$-symmetry,
i.e.~type I--IV models. In the following we define three reference
scenarios of this type in order to compare and discuss differences
between the codes for the evaluation of cross sections and decay
rates. Please note that these scenarios have been chosen arbitrarily,
and should not be interpreted as benchmark scenarios for 2HDM
searches. The input parameters for these scenarios are given in
Table~\ref{tab:param}. For the numerical results presented below we have
adopted values for the SM parameters according to Appendix~\ref{app:SM}. 
Of course, the quoted reference values for the 2HDM cross sections and branching ratios 
depend sensitively on these parameter settings.

\begin{table}
\centering
\begin{tabular}{lccc}
\hline
Parameter                       & Scenario A & Scenario B & Scenario C \\
\hline\hline
\multicolumn{4}{c}{Common parameters}\\\hline
Type                            & I             & II            & II\\
$M_h$ (GeV)                     & $125$   & $125$       & $125$ \\
$M_H$ (GeV)             & $300$   & $300$   & $400$ \\
$M_A$ (GeV)             & $330$   & $270$       & $500$ \\
$M_{H^\pm}$ (GeV)       & $230$    & $335$      & $550$ \\
$M_{12}^2$ (GeV$^2$)& $25600$ & $1798$  & $15800$\\
$\tan\beta$                     & $1.5$   & $50$        & $10$\\
\hline\hline
\multicolumn{4}{c}{\tt 2HDMC}\\\hline
$\sin(\beta-\alpha)$& $0.901314$ & $0.999001$ & $0.999$ \\
$\lambda_6$ & 0 & 0 & 0\\
$\lambda_7$ & 0 & 0 & 0\\
\hline\hline
\multicolumn{4}{c}{\tt HDECAY}\\\hline
$\alpha$                        & $-0.14$ & $0.0247$& $-0.0549436$ \\
\hline
\end{tabular}
\caption{Values of the input parameters for the 2HDM reference scenarios.}
\label{tab:param}
\end{table}


\subsection{Cross sections}

Tables\,\ref{tab:gghA}--\ref{tab:gghC} show the cross sections for
scenarios A, B, and C as evaluated with {\tt SusHi} and {\tt HIGLU}.
The columns denoted ``\%'' show the percental deviation between the {\tt
  SusHi} and the {\tt HIGLU} result, defined as
\begin{equation}
\begin{split}
\Delta \equiv 2\left|\frac{\sigma_\text{\tt SusHi} - \sigma_\text{\tt
    HIGLU}}{\sigma_\text{\tt SusHi} + \sigma_\text{\tt HIGLU}}\right|\,.
\end{split}
\end{equation}
We have checked that the agreement between these two programs in all
cases is better than 0.1\% if the same set of corrections is taken into
account.  The numerical differences observed in the tables for the
CP-even Higgs bosons are due to the electroweak corrections taken into
account by {\tt SusHi}, as described in Section\,\ref{sec:xsec2hdm}. For
the CP-odd Higgs boson, these corrections are absent due to the
vanishing $AVV$ coupling at tree level.

The numbers listed in the tables correspond to our recommendations for
the central values of the cross section, which are obtained with the
central MSTW2008 parton density sets. For gluon fusion, the
renormalization and factorization scale is set to
$\mu_F=\mu_R=M_\phi/2$, where $\phi\in\{h,H,A\}$. Theoretical
uncertainties should be determined by varying $\mu_F$ and $\mu_R$
independently within a factor of two around $M_\phi/2$. For $b\bar
b\phi$ production in the 5FS, we use $\mu_R=M_\phi$ and
$\mu_F=M_\phi/4$\,\cite{Maltoni:2003pn,Harlander:2003ai} as central
scales.  The scale uncertainty can be obtained as suggested in
Ref.\,\cite{Harlander:2003ai,Dittmaier:2011ti} by varying $\mu_F$ within
the interval $[0.1,0.7]M_\phi$ and $\mu_R$ within $[1/5,5]M_\phi$. PDF
uncertainties for the 5FS can be estimated following the recommendation
of the PDF4LHC group\,\cite{pdf4lhc}.  In the 4FS numbers of
Ref.\,\cite{4fsgrids}, it is\footnote{In Ref.\,\cite{Dittmaier:2011ti},
  it is stated by mistake that the Yukawa coupling is renormalized at
  $\mu_R=M_\phi$.} $\mu_F=\mu_R=M_\phi/4$.  The uncertainties have been
obtained by varying both $\mu_F$ and $\mu_R$ within the interval
$[M_\phi/8,M_\phi/2]$.

In addition, the uncertainty due to the undetermined renormalization
scheme for the bottom Yukawa coupling should be taken into
account. Clearly, its impact on the 2HDM cross section prediction can be
much more severe than in the SM, in particular in type\,II models with
large $\tan\beta$. Consider, for example, the results for reference
scenario\,B at 14\,TeV. The total cross sections for the three Higgs
bosons can be decomposed as
\begin{equation}
\begin{split}
\sigma(gg\to h) &= 55.55\,\left( 0.93 + 0.06 + 0.01 \right)\text{pb}
+ 2.23\,\text{pb} \,,\\
\sigma(gg\to H) &= 8.837\,\left( 0.00 - 0.03 + 1.03 \right)\text{pb}
+ 0.007\,\text{pb} \,,\\
\sigma(gg\to A) &= 16.49\,\left( 0.00 - 0.03 + 1.03 \right)\text{pb} \,,
\end{split}
\end{equation}
where the first and the third number in brackets correspond to the pure
top and bottom quark contribution, respectively, and the second number
to the interference term. The last term, which is added to the total result, 
corresponds to the electroweak corrections. For the heavy Higgs bosons $H,A$ the bottom
loops provide by far the dominant contribution in scenario B. The
associated uncertainties are sizable due to the choice of the bottom
mass inside the loops and the bottom Yukawa coupling and should be taken
over from the corresponding analyses within the MSSM working group, see
\cite{4fsgrids}.

\begin{table}[h]
\begin{center}
\input{tables/table-ggh-A}
\caption[]{\label{tab:gghA} Gluon fusion cross sections for scenario\,A.}
\end{center}
\end{table}

\begin{table}[h]
\begin{center}
\input{tables/table-ggh-B}
\caption[]{\label{tab:gghB} Gluon fusion cross sections for scenario\,B.}
\end{center}
\end{table}
\clearpage

\begin{table}[h]
\begin{center}
\input{tables/table-ggh-C}
\caption[]{\label{tab:gghC} Gluon fusion cross sections for scenario\,C.}
\end{center}
\end{table}

Tables \ref{tab:bbhA}--\ref{tab:bbhC} display the cross section for
associated $b\bar b\phi$ production in the three reference scenarios
considered here. They list the total inclusive cross
section as obtained in the 5FS, which can be calculated along with the
$gg\to \phi$ cross section with {\tt SusHi}; in addition, the 4FS result
as obtained from the grids of the LHCHXSWG is displayed, see
\cite{4fsgrids}. The third number, labeled ``matched'' in the table,
corresponds to the combined result according to the Santander matching
prescription:
\begin{equation}
\begin{split}
\sigma_\text{matched} = \frac{\sigma^\text{4FS} +
  w\sigma^\text{5FS}}{1+w}\,,
\end{split}
\end{equation}
where the weight function $w\equiv \ln(m_\phi/m_b)-2$ is chosen such
as to account for the excellent agreement of the two predictions at
$m_\phi=100$\,GeV, i.e., $w|_{m_\phi=100\text{GeV}} \approx 1$. We
also recommend to follow Ref.\,\cite{santander} in determining the theoretical
uncertainties. 

\begin{table}[h]
\begin{center}
\input{tables/table-bbh-A}
\caption[]{\label{tab:bbhA} Associated $b\bar b\phi$ cross sections for
  scenario\,A. Note that the numbers for the heavy and the CP-odd Higgs
  are rescaled by a factor $10^{4}$ and $10^{3}$, respectively.}
\end{center}
\end{table}
\begin{table}[h]
\begin{center}
\input{tables/table-bbh-B}
\caption[]{\label{tab:bbhB} Associated $b\bar b\phi$ cross sections for
  scenario\,B.}
\end{center}
\end{table}
\begin{table}[h]
\begin{center}
\input{tables/table-bbh-C}
\caption[]{\label{tab:bbhC} Associated $b\bar b\phi$ cross sections for
  scenario\,C.}
\end{center}
\end{table}

\clearpage
\newpage

\subsection{Branching ratios}


Tables~\ref{tab:compA}--\ref{tab:compC} show numerical results on the branching ratios and partial decay widths (in GeV) for the neutral Higgs bosons in scenarios A--C. In each table, the
first two columns show the results obtained with {\tt 2HDMC-1.6.2}, the next two columns the same quantities computed with {\tt HDECAY-6.00}, and the final column gives, for each decay, the ratio of calculated widths,
$\Gamma_{\rm 2HDMC}/\Gamma_{\rm HDECAY}$. Since the results from comparing {\tt 2HDMC} and {\tt HDECAY} are qualitatively very similar for all three scenarios, we discuss Tables~\ref{tab:compA}--\ref{tab:compC} together.

For decays calculated at leading order, the relative differences in the partial widths are at the level of $10^{-3}$ (or smaller) for all three neutral Higgs bosons. This includes the decays into lepton pairs, doube off-shell vector bosons, cascade Higgs decays, and mixed Higgs/vector boson final states, which are all in perfect agreement. 

For the decays of the lightest Higgs boson, $h$, the largest relative difference is at the level of $2\%$, for the decay into $c\bar{c}$. Relative differences at the level of $1$--$2\%$ are also observed for the $s\bar{s}$ final state, with a smaller deviation ($0.8\%$) in the numerically most important result for $h\to b\bar{b}$. Similar numbers are found for the decays of the heavier Higgs bosons, $H$ and $A$, into light quarks. It can be seen that these small differences are generic, and we have verified explicitly that they are due to small differences in the values of the running quark masses (which are evaluated at the scale of the Higgs mass for these decays).

For the phenomenologically important decay $h\to \gamma\gamma$, the level of agreement is better than $1\%$, whereas for $h\to gg$ the difference corresponds to around $1.3\%$. The same decays for the heavier Higgs boson show typically slightly larger deviations in the results, at the level of $2$--$3\%$. For these loop-induced decays, a residual difference of this magnitude is to be expected due to the different implementation of QCD corrections in the two codes (for details, see above). We have verified that this is indeed the source of the observed differences. One case where the full mass dependence in the QCD corrections becomes important is for the decays $H/A\to \gamma\gamma$ close to, or above, the $t\bar{t}$ threshold (cf.~Scenarios A and C). Here, the full $\alpha_s$ corrections (as implemented in {\tt HDECAY}) lead to a strong enhancement of this decay. However, since the corresponding branching ratios are very small, below $10^{-4}$, this difference does not affect the total width (and not the branching ratios). 

In all three reference scenarios, the calculated total widths of the neutral Higgs bosons agree at the sub-percent level between {\tt 2HDMC} and {\tt HDECAY}, which demonstrates consistent results on the 2HDM branching ratios at a sufficient level of precision.

\begin{table}
\vspace*{-1.5cm}

\centering
\input{tables/table-br-A}
\caption{Numerical comparison between neutral Higgs  branching ratios and decay widths
 calculated with {\tt 2HDMC} and {\tt HDECAY} for
  reference scenario A.}
\label{tab:compA}
\end{table}

\begin{table}
\centering
\input{tables/table-br-B}
\caption{Numerical comparison between neutral Higgs  branching ratios and decay widths calculated with {\tt 2HDMC} and {\tt HDECAY} for
  reference scenario B.}
\label{tab:compB}
\end{table}

\begin{table}
\vspace*{-0.5cm}

\centering
\input{tables/table-br-C}
\caption{Numerical comparison between neutral Higgs  branching ratios and decay widths calculated with {\tt 2HDMC} and {\tt HDECAY} for
  reference scenario C.}
\label{tab:compC}
\end{table}
\clearpage



\section{Conclusions}
We have described the current status of theoretical predictions for the
gluon fusion and associated $b\bar b\phi$ Higgs production processes as
well as the Higgs branching ratios within the 2HDM, resulting in an
interim recommendation of the LHC Higgs Cross Section Working Group for the
experimental collaborations. The theoretical
tools recommended to evaluate these quantities are {\tt HIGLU} and {\tt SusHi}
for the total inclusive cross sections, and {\tt 2HDMC} and {\tt HDECAY}
for the decay rates and branching ratios. These programs are publically
available from the following URLs:

{\tt
\begin{tabular}{ll}
SusHi: & \url{http://sushi.hepforge.org/}\\
2HDMC: & \url{http://2hdmc.hepforge.org/}\\
HIGLU: & \url{http://people.web.psi.ch/spira/higlu/}\\
HDECAY:& \url{http://people.web.psi.ch/spira/hdecay/}
\end{tabular}
}

Please always use the latest versions of these programs; the results
presented in this note have been obtained with
{\tt 2HDMC-1.6.2}, {\tt HDECAY-6.00}, {\tt HIGLU-4.00}, and {\tt
  SusHi-1.1.1}. Earlier versions of these programs must not be used to
obtain viable 2HDM results.

Currently, {\tt SusHi} and {\tt 2HDMC} can be linked to each other, and
run with a single input file. In addition a similar approach
has been adopted for {\tt HIGLU} and {\tt HDECAY} which are linked but
work with two input files, i.e.~the usual input file of {\tt HDECAY} and
an additional one for {\tt HIGLU} that specifies properties of the
production cross sections to be calculated beyond the input values of
{\tt HDECAY}.

As pointed out earlier, these recommendations are based on currently
available results which are implemented in public tools.  We hope to update
them as more advanced calculations or tools become available.

\section*{Acknowledgements}
We thank the members of the Heavy Higgs Group of the LHC Higgs cross section working group for discussions. In particular we are grateful to Nikolaos Rompotis for useful comments and suggestions. M.M.~thanks Rui Santos for helpful discussions. R.H. is supported by BMBF (grant no.~05H12PXE) and DFG (grant no.~HA 2990/6-1).  M.M.~is
supported by the DFG SFB/TR9~``Computational Particle Physics''.
J.R.~is supported in part by the Swedish Research Council under
contracts 621-2011-5333 and 621-2013-4287. O.S.~is supported by the
Swedish Research Council (VR) through the Oskar Klein Centre.



\clearpage
\newpage
\appendix
\section{SM parameters}
\label{app:SM}
For the numerical comparisons presented in this note, the following set
of SM input parameters has been adopted. The numbers are given in the
input style of {\tt HDECAY}, but the same parameter values were used
with {\tt 2HDMC}.
\begin{verbatim}
ALS(MZ)  = 0.119D0
MSBAR(2) = 0.100D0
MC       = 1.42D0
MB       = 4.75D0
MT       = 172.5D0
MTAU     = 1.77684D0
MMUON    = 0.105658367D0
1/ALPHA  = 137.0359997D0
GF       = 1.16637D-5
GAMW     = 2.08856D0
GAMZ     = 2.49581D0
MZ       = 91.15349D0
MW       = 80.36951D0
VTB      = 0.9991D0
VTS      = 0.0404D0
VTD      = 0.00867D0
VCB      = 0.0412D0
VCS      = 0.97344D0
VCD      = 0.22520D0
VUB      = 0.00351D0
VUS      = 0.22534D0
VUD      = 0.97427D0
\end{verbatim}

In both codes, {\tt 2HDMC} and {\tt HDECAY}, three-loop running of the
strong coupling has been used. Matching between the quark flavor
thresholds was performed at the pole masses, taking into account
threshold corrections to $\mathcal{O}(\alpha_s^2)$. The running quark
masses have been computed at one loop in the $\overline{\mathrm{MS}}$
scheme from the input pole masses given above, and running was then
performed at three loops with matching at the pole masses. Threshold
corrections have not been included for the running quark masses. Note in
addition that the $W$ and $Z$ masses correspond to the pole masses in
the complex-mass scheme, and that the top and bottom quark masses correspond to the
standard choices of the SM input parameters of the LHC Higgs Cross
Sections Working Group \cite{Dittmaier:2011ti}.


\end{document}

%% file: tables/table-global-k-8.tex
\begin{tabular}{|l|ccc|c|} \hline
$M_H$ [GeV] & $\sigma_{tt}$ [pb] & $\sigma_{tb}$ [pb] & $\sigma_{bb}$ [pb]
& $\sigma_{tot}$ [pb] \\ \hline \hline
& (LO) & (LO) & (LO) & (LO) \\ \hline \hline
125.0 &    10.32  &    -1.241   &   0.1131    &  9.188 \\
125.5 &    10.23  &    -1.225   &   0.1108    &  9.116 \\
126.0 &    10.15  &    -1.210   &   0.1085    &  9.044 \\
150.0 &     7.012 &    -0.6770  &   0.04330   &  6.379 \\
200.0 &     3.788 &    -0.2465  &   0.009013  &  3.551 \\
250.0 &     2.395 &    -0.1093  &   0.002540  &  2.288 \\
300.0 &     1.748 &    -0.05662 &   0.0008719 &  1.692 \\
\hline \hline
& (NLO) & (NLO) & (NLO) & (NLO) \\ \hline \hline
125.0 & 17.32  & -1.206   & 0.1357   & 16.25 \\
125.5 & 17.18  & -1.192   & 0.1330   & 16.12 \\
126.0 & 17.04  & -1.178   & 0.1303   & 15.99 \\
150.0 & 11.90  & -0.6920  & 0.05293  & 11.26 \\
200.0 &  6.518 & -0.2694  & 0.01133  & 6.260 \\
250.0 &  4.159 & -0.1250  & 0.003257 & 4.037 \\
300.0 &  3.061 & -0.06705 & 0.001135 & 2.995 \\
\hline \hline
& (NNLO) & (NLO) & (NLO) & (mixed) \\ \hline \hline
125.0 & 20.18  & -1.206   & 0.1357   & 19.11 \\
125.5 & 20.02  & -1.192   & 0.1330   & 18.96 \\
126.0 & 19.85  & -1.178   & 0.1303   & 18.81 \\
150.0 & 13.78  & -0.6920  & 0.05293  & 13.14 \\
200.0 &  7.490 & -0.2694  & 0.01133  &  7.231 \\
250.0 &  4.754 & -0.1250  & 0.003257 &  4.633 \\
300.0 &  3.485 & -0.06705 & 0.001135 &  3.419 \\ \hline
\end{tabular}

%% file: tables/table-global-k-14.tex
\begin{tabular}{|l|ccc|c|} \hline
$M_H$ [GeV] & $\sigma_{tt}$ [pb] & $\sigma_{tb}$ [pb] & $\sigma_{bb}$ [pb]
& $\sigma_{tot}$ [pb] \\ \hline \hline
& (LO) & (LO) & (LO) & (LO) \\ \hline \hline
125.0 & 27.92  & -3.360  & 0.3061   & 24.87 \\
125.5 & 27.72  & -3.320  & 0.3001   & 24.70 \\
126.0 & 27.51  & -3.281  & 0.2942   & 24.52 \\
150.0 & 19.85  & -1.916  & 0.1226   & 18.05 \\
200.0 & 11.60  & -0.7547 & 0.02760  & 10.87 \\
250.0 &  7.860 & -0.3588 & 0.008335 &  7.509 \\
300.0 &  6.112 & -0.1980 & 0.003048 &  5.917 \\ \hline \hline
& (NLO) & (NLO) & (NLO) & (NLO) \\ \hline \hline
125.0 & 44.05 & -2.799  & 0.3302   & 41.58 \\
125.5 & 43.74 & -2.770  & 0.3239   & 41.29 \\
126.0 & 43.43 & -2.741  & 0.3177   & 41.01 \\
150.0 & 31.72 & -1.695  & 0.1351   & 30.16 \\
200.0 & 18.88 & -0.7236 & 0.03144  & 18.19 \\
250.0 & 12.96 & -0.3634 & 0.009741 & 12.60 \\
300.0 & 10.18 & -0.2094 & 0.003635 &  9.979 \\
\hline \hline
& (NNLO) & (NLO) & (NLO) & (mixed) \\ \hline \hline
125.0 & 51.61 & -2.799  & 0.3302   & 49.14 \\
125.5 & 51.24 & -2.770  & 0.3239   & 48.79 \\
126.0 & 50.87 & -2.741  & 0.3177   & 48.44 \\
150.0 & 36.91 & -1.695  & 0.1351   & 35.35 \\
200.0 & 21.77 & -0.7236 & 0.03144  & 21.08 \\
250.0 & 14.85 & -0.3634 & 0.009741 & 14.50 \\
300.0 & 11.62 & -0.2094 & 0.003635 & 11.42 \\ \hline
\end{tabular}

%% file: tables/table-ggh-A.tex
\begin{tabular}{|c|ccc|ccc|ccc|}
\hline$\sqrt{s}$ & \multicolumn{3}{|c|}{$\sigma(gg\to h)$/pb} &\multicolumn{3}{|c|}{ $\sigma(gg\to H)$/pb} &\multicolumn{3}{|c|}{ $\sigma(gg\to A)$/pb} \\
  /TeV &{\tt SusHi} & {\tt HIGLU} & $\%$ &{\tt SusHi} & {\tt HIGLU} & $\%$ &{\tt SusHi} & {\tt HIGLU} & $\%$ \\\hline
$ 7$ & $  22.00$ & $  21.25$ & $  3.5$ & $  0.07199$ & $  0.06996$ & $  2.9$ & $  4.061$ & $  4.063$ & $  0.05$\\ 
 $ 8$ & $  28.03$ & $  27.07$ & $  3.5$ & $  0.09897$ & $  0.09617$ & $  2.9$ & $  5.639$ & $  5.642$ & $  0.06$\\ 
 $ 13$ & $  63.96$ & $  61.79$ & $  3.5$ & $  0.2846$ & $  0.2766$ & $  2.9$ & $  16.69$ & $  16.70$ & $  0.04$\\ 
 $ 14$ & $  72.05$ & $  69.60$ & $  3.5$ & $  0.3305$ & $  0.3212$ & $  2.8$ & $  19.45$ & $  19.46$ & $  0.04$\\ 
 \hline
\end{tabular}

%% file: tables/table-ggh-B.tex
\begin{tabular}{|c|ccc|ccc|ccc|}
\hline$\sqrt{s}$ & \multicolumn{3}{|c|}{$\sigma(gg\to h)$/pb} &\multicolumn{3}{|c|}{ $\sigma(gg\to H)$/pb} &\multicolumn{3}{|c|}{ $\sigma(gg\to A)$/pb} \\
  /TeV &{\tt SusHi} & {\tt HIGLU} & $\%$ &{\tt SusHi} & {\tt HIGLU} & $\%$ &{\tt SusHi} & {\tt HIGLU} & $\%$ \\\hline
$ 7$ & $  17.91$ & $  17.22$ & $  3.9$ & $  2.025$ & $  2.023$ & $  0.1$ & $  3.979$ & $  3.978$ & $  0.03$\\ 
 $ 8$ & $  22.76$ & $  21.88$ & $  3.9$ & $  2.759$ & $  2.756$ & $  0.1$ & $  5.355$ & $  5.354$ & $  0.03$\\ 
 $ 13$ & $  51.37$ & $  49.40$ & $  3.9$ & $  7.660$ & $  7.652$ & $  0.1$ & $  14.35$ & $  14.34$ & $  0.05$\\ 
 $ 14$ & $  57.78$ & $  55.56$ & $  3.9$ & $  8.844$ & $  8.835$ & $  0.1$ & $  16.49$ & $  16.48$ & $  0.06$\\ 
 \hline
\end{tabular}

%% file: tables/table-ggh-C.tex
\begin{tabular}{|c|ccc|ccc|ccc|}
\hline$\sqrt{s}$ & \multicolumn{3}{|c|}{$\sigma(gg\to h)$/pb} &\multicolumn{3}{|c|}{ $\sigma(gg\to H)$/pb} &\multicolumn{3}{|c|}{ $\sigma(gg\to A)$/pb} \\
  /TeV &{\tt SusHi} & {\tt HIGLU} & $\%$ &{\tt SusHi} & {\tt HIGLU} & $\%$ &{\tt SusHi} & {\tt HIGLU} & $\%$ \\\hline
$ 7$ & $  16.17$ & $  15.47$ & $  4.4$ & $  0.02707$ & $  0.02702$ & $  0.2$ & $  0.02421$ & $  0.02422$ & $  0.05$\\ 
 $ 8$ & $  20.59$ & $  19.70$ & $  4.4$ & $  0.03821$ & $  0.03814$ & $  0.2$ & $  0.03575$ & $  0.03576$ & $  0.03$\\ 
 $ 13$ & $  46.88$ & $  44.86$ & $  4.4$ & $  0.1179$ & $  0.1177$ & $  0.2$ & $  0.1269$ & $  0.1269$ & $  0.01$\\ 
 $ 14$ & $  52.80$ & $  50.52$ & $  4.4$ & $  0.1380$ & $  0.1377$ & $  0.2$ & $  0.1513$ & $  0.1514$ & $  0.07$\\ 
 \hline
\end{tabular}

%% file: tables/table-bbh-A.tex
\begin{tabular}{|c|ccc|ccc|ccc|}
\hline$\sqrt{s}$ & \multicolumn{3}{|c|}{$\sigma(b\bar b\to h)$/pb}
&\multicolumn{3}{|c|}{ $\sigma(b\bar b\to H)$/pb $\cdot 10^4$}
&\multicolumn{3}{|c|}{ $\sigma(b\bar b\to A)$/pb $\cdot 10^3$} \\ /TeV &
5FS & 4FS & matched & 5FS & 4FS & matched& 5FS & 4FS & matched \\\hline
$ 7$ & $ 0.239$ & $ 0.221$ & $ 0.231$ & $ 1.45$ & $ 1.18$ & $ 1.37$ & $
1.49$ & $ 1.19$ & $ 1.39$\\ $ 8$ & $ 0.310$ & $ 0.293$ & $ 0.302$ & $
2.07$ & $ 1.69$ & $ 1.95$ & $ 2.15$ & $ 1.74$ & $ 2.02$\\ $ 13$ & $
0.743$ & $ 0.727$ & $ 0.736$ & $ 6.56$ & $ 5.58$ & $ 6.25$ & $ 7.08$ & $
5.96$ & $ 6.74$\\ $ 14$ & $ 0.842$ & $ 0.822$ & $ 0.833$ & $ 7.70$ & $
6.68$ & $ 7.38$ & $ 8.36$ & $ 7.13$ & $ 7.98$\\ \hline
\end{tabular}

%% file: tables/table-bbh-B.tex
\begin{tabular}{|c|ccc|ccc|ccc|}
\hline$\sqrt{s}$ & \multicolumn{3}{|c|}{$\sigma(b\bar b\to h)$/pb} &\multicolumn{3}{|c|}{ $\sigma(b\bar b\to H)$/pb} &\multicolumn{3}{|c|}{ $\sigma(b\bar b\to A)$/pb} \\
  /TeV & 5FS & 4FS & matched & 5FS & 4FS & matched& 5FS & 4FS & matched \\\hline
$ 7$ & $  0.257$ & $  0.238$ & $  0.249$ & $  12.9$ & $  10.5$ & $  12.1$ & $  20.6$ & $  17.0$ & $  19.4$\\ 
 $ 8$ & $  0.334$ & $  0.314$ & $  0.325$ & $  18.4$ & $  15.1$ & $  17.3$ & $  28.9$ & $  24.2$ & $  27.4$\\ 
 $ 13$ & $  0.801$ & $  0.784$ & $  0.793$ & $  58.3$ & $  49.5$ & $  55.5$ & $  88.0$ & $  76.5$ & $  84.2$\\ 
 $ 14$ & $  0.907$ & $  0.885$ & $  0.897$ & $  68.5$ & $  59.4$ & $  65.6$ & $ 103$ & $  89.9$ & $  98.5$\\ 
 \hline
\end{tabular}

%% file: tables/table-bbh-C.tex
\begin{tabular}{|c|ccc|ccc|ccc|}
\hline$\sqrt{s}$ & \multicolumn{3}{|c|}{$\sigma(b\bar b\to h)$/pb} &\multicolumn{3}{|c|}{ $\sigma(b\bar b\to H)$/pb} &\multicolumn{3}{|c|}{ $\sigma(b\bar b\to A)$/pb} \\
  /TeV & 5FS & 4FS & matched & 5FS & 4FS & matched& 5FS & 4FS & matched \\\hline
$ 7$ & $  0.0513$ & $  0.0476$ & $  0.0497$ & $  0.135$ & $  0.104$ & $  0.126$ & $  0.0430$ & $  0.0318$ & $  0.0399$\\ 
 $ 8$ & $  0.0667$ & $  0.0628$ & $  0.0650$ & $  0.200$ & $  0.156$ & $  0.187$ & $  0.0666$ & $  0.0500$ & $  0.0621$\\ 
 $ 13$ & $  0.160$ & $  0.156$ & $  0.158$ & $  0.723$ & $  0.594$ & $  0.686$ & $  0.271$ & $  0.215$ & $  0.256$\\ 
 $ 14$ & $  0.181$ & $  0.177$ & $  0.179$ & $  0.863$ & $  0.708$ & $  0.818$ & $  0.329$ & $  0.263$ & $  0.311$\\ 
 \hline
\end{tabular}

%% file: tables/table-br-A.tex
\begin{tabular}{rl|cc|cc|c}
\hline
& &  \multicolumn{2}{c|}{\tt 2HDMC} & \multicolumn{2}{c|}{\tt HDECAY}\\
& &  $\mathrm{BR}$ & $\Gamma$ (GeV) & $\mathrm{BR}$ & $\Gamma$ (GeV) & $\Gamma_{\mathrm{2H}}/\Gamma_{\mathrm{HD}}$\\
\hline
$h \to$ & $b\bar{b}$& $0.6812$ & $3.790\times 10^{-3}$ & $0.6827$& $3.820\times 10^{-3}$ & $0.992$\\
&$\tau^+\tau^-$   & $6.587\times 10^{-2}$ & $3.664\times 10^{-4}$ & $6.548\times 10^{-2}$ & $3.664\times 10^{-4}$& $1.000$\\
&$\mu^+\mu^-$   & $2.332\times 10^{-4}$ & $1.297\times 10^{-6}$ & $2.318\times 10^{-4}$ & $1.297\times 10^{-6}$ & $1.000$\\
&$s\bar{s}$   & $2.484\times 10^{-4}$ & $1.382\times 10^{-6}$ & $2.503\times 10^{-4}$ & $1.400\times 10^{-6}$ & $0.987$\\
&$c\bar{c}$   & $3.059\times 10^{-2}$ & $1.701\times 10^{-4}$ & $2.976\times 10^{-2}$ & $1.665\times 10^{-4}$ & $1.022$\\
&$gg$  & $8.110\times 10^{-2}$ & $4.511\times 10^{-4}$ & $8.166\times 10^{-2}$& $4.569\times 10^{-4}$ & $0.987$\\
&$\gamma\gamma$   & $1.130\times 10^{-3}$ & $6.284\times 10^{-6}$ & $1.117\times 10^{-3}$ & $6.250\times 10^{-6}$ & $1.006$\\
&$Z\gamma$   & $8.728\times 10^{-4}$ & $4.855\times 10^{-6}$ & $8.677\times 10^{-4}$ & $4.855\times 10^{-6}$ & $1.000$\\
&$W^+W^-$   & $0.1233$ & $6.859\times 10^{-4}$ & $0.1226$ & $6.860\times 10^{-4}$ & $1.000$\\
&$ZZ$   & $1.540\times 10^{-2}$ & $8.569\times 10^{-5}$ & $1.531\times 10^{-2}$ & $8.566\times 10^{-5}$ & $1.000$\\
\hline
\multicolumn{2}{c|}{Total width}& & $\bf{5.563\times 10^{-3}}$ & & $\mathbf{5.595\times 10^{-3}}$ & $\mathbf{0.994}$  \\
\hline\hline
$H\to$ & $b\bar{b}$ & $8.492\times 10^{-5}$& $1.536\times 10^{-4}$& $8.526\times 10^{-5}$& $1.542\times 10^{-4}$ & $0.996$\\
&$\tau^+\tau^-$     & $9.667\times 10^{-6}$& $1.748\times 10^{-5}$& $9.667\times 10^{-6}$& $1.748\times 10^{-5}$& $1.000$\\
&$\mu^+\mu^-$   & $3.419\times 10^{-8}$ & $6.182\times 10^{-8}$ & $3.419\times 10^{-8}$ & $6.183\times 10^{-8}$& $1.000$\\
&$s\bar{s}$   & $3.070\times 10^{-8}$ & $5.552\times 10^{-8}$ & $3.115\times 10^{-7}$& $5.636\times 10^{-8}$& $0.985$\\
&$c\bar{c}$   & $3.787\times 10^{-6}$ & $6.848\times 10^{-6}$ & $3.706\times 10^{-6}$ & $6.706\times 10^{-6}$ & $1.021$\\
&$t\bar{t}$& $5.976\times 10^{-6}$ & $1.081\times 10^{-5}$ & $5.986\times 10^{-6}$ & $1.082\times 10^{-5}$ & $0.998$\\
&$gg$   & $8.382\times 10^{-5}$ & $1.516\times 10^{-4}$ & $8.669\times 10^{-5}$ & $1.568\times 10^{-4}$ & $0.967$\\
&$\gamma\gamma$   & $1.642\times 10^{-5}$ & $2.969\times 10^{-5}$ & $1.653\times 10^{-5}$ & $2.989\times 10^{-5}$ & $0.993$\\
&$Z\gamma$   & $5.300\times 10^{-5}$ & $9.584\times 10^{-5}$ & $5.300\times 10^{-5}$ & $9.584\times 10^{-5}$ & $1.000$\\
&$W^+W^-$    & $0.5872$ & $1.062$ & $0.5872$& $1.062$ & $1.000$\\
&$ZZ$   & $0.2606$ & $0.4713$ & $0.2606$ & $0.4712$ & $1.000$\\
&$hh$   & $0.1493$& $0.2699$ & $0.1493$ & $0.2700$ & $1.000$\\
&$W^\pm H^\mp$ & $2.658\times 10^{-3}$& $4.806\times 10^{-3}$ & $2.663\times 10^{-3}$ &  $4.815\times 10^{-3}$& $0.998$\\
\hline
\multicolumn{2}{c|}{Total width} &  & $\mathbf{1.808}$ & & $\mathbf{1.808}$  & $\mathbf{1.000}$  \\
\hline\hline
$A\to$ & $b\bar{b}$ & $1.568\times 10^{-3}$& $2.564\times 10^{-3}$& $1.573\times 10^{-3}$& $2.572\times 10^{-3}$ & $0.997$\\
&$\tau^+\tau^-$     & $1.859\times 10^{-4}$& $3.039\times 10^{-4}$& $1.858\times 10^{-4}$& $3.038\times 10^{-4}$& $1.000$\\
&$\mu^+\mu^-$   & $6.573\times 10^{-7}$ & $1.075\times 10^{-6}$ & $6.571\times 10^{-7}$ & $1.075\times 10^{-6}$& $1.000$\\
&$s\bar{s}$   & $5.385\times 10^{-7}$ & $8.804\times 10^{-7}$ & $5.466\times 10^{-7}$& $8.939\times 10^{-7}$& $0.985$\\
&$c\bar{c}$   & $7.219\times 10^{-5}$ & $1.180\times 10^{-4}$ & $7.067\times 10^{-5}$ & $1.156\times 10^{-4}$ & $1.021$\\
&$t\bar{t}$   & $1.395\times 10^{-2}$ & $2.280\times 10^{-2}$ & $1.399\times 10^{-2}$ & $2.288\times 10^{-2}$ & $0.997$\\
&$gg$   & $8.874\times 10^{-3}$ & $1.451\times 10^{-2}$ & $9.060\times 10^{-3}$ & $1.482\times 10^{-2}$ & $0.979$\\
&$\gamma\gamma$   & $2.380\times 10^{-5}$ & $3.891\times 10^{-5}$ & $3.155\times 10^{-5}$ & $5.159\times 10^{-5}$ & $0.754$\\
&$Z\gamma$   & $5.725\times 10^{-6}$ & $9.360\times 10^{-6}$ & $5.724\times 10^{-6}$ & $9.361\times 10^{-6}$ & $1.000$\\
&$Zh$   & $0.5747$ & $0.9396$ & $0.5746$& $0.9397$ & $1.000$\\
&$ZH$   & $2.221\times 10^{-6}$ & $9.852\times 10^{-6}$ & $6.029\times 10^{-6}$ & $9.859\times 10^{-6}$ & $0.999$\\
&$W^\pm H^\mp$ & $0.4006$& $0.6550$   & $0.4005$ & $0.6550$ & $1.000$\\
\hline
\multicolumn{2}{c|}{Total width} &  & $\mathbf{1.635}$ & & $\mathbf{1.635}$  & $\mathbf{1.000}$  \\\hline\hline
\end{tabular}

%% file: tables/table-br-B.tex
\begin{tabular}{rl|cc|cc|c}
\hline
& &  \multicolumn{2}{c|}{\tt 2HDMC} & \multicolumn{2}{c|}{\tt HDECAY}\\
& &  $\mathrm{BR}$ & $\Gamma$ (GeV) & $\mathrm{BR}$ & $\Gamma$ (GeV) & $\Gamma_{\mathrm{2H}}/\Gamma_{\mathrm{HD}}$\\
\hline
$h \to$ & $b\bar{b}$& $0.6794$ & $4.018\times 10^{-3}$ & $0.6809$& $4.051\times 10^{-3}$ & $0.992$\\
&$\tau^+\tau^-$   & $6.672\times 10^{-2}$ & $3.946\times 10^{-4}$ & $6.633\times 10^{-2}$ & $3.947\times 10^{-4}$& $1.000$\\
&$\mu^+\mu^-$   & $2.362\times 10^{-4}$ & $1.397\times 10^{-6}$ & $2.348\times 10^{-4}$ & $1.397\times 10^{-6}$ & $1.000$\\
&$s\bar{s}$   & $2.384\times 10^{-4}$ & $1.410\times 10^{-6}$ & $2.402\times 10^{-6}$ & $1.429\times 10^{-6}$ & $0.987$\\
&$c\bar{c}$   & $2.031\times 10^{-2}$ & $1.201\times 10^{-4}$ & $1.975\times 10^{-2}$ & $1.175\times 10^{-4}$ & $1.022$\\
&$gg$  & $7.043\times 10^{-2}$ & $4.166\times 10^{-4}$ & $7.089\times 10^{-2}$ & $4.218\times 10^{-4}$ & $0.988$\\
&$\gamma\gamma$   & $1.391\times 10^{-3}$ & $8.229\times 10^{-6}$ & $1.375\times 10^{-3}$ & $8.181\times 10^{-6}$ & $1.006$\\
&$Z\gamma$   & $1.013\times 10^{-3}$ & $5.994\times 10^{-6}$ & $1.003\times 10^{-3}$ & $5.968\times 10^{-6}$ & $1.004$\\
&$W^+W^-$& $0.1425$ & $8.427\times 10^{-4}$ & $0.1416$ & $8.425\times 10^{-4}$ & $1.000$\\
&$ZZ$   & $1.780\times 10^{-2}$ & $1.053\times 10^{-4}$ & $1.769\times 10^{-2}$ & $1.053\times 10^{-4}$ & $1.000$\\
\hline
\multicolumn{2}{c|}{Total width}& & $\bf{5.915\times 10^{-3}}$ & & $\mathbf{5.950\times 10^{-3}}$ & $\mathbf{0.994}$  \\
\hline\hline
$H\to$ & $b\bar{b}$ & $0.8947$ & $13.52$ & $0.8950$& $13.58$& $0.996$\\
&$\tau^+\tau^-$     & $0.1028$ & $1.553$ & $0.1024$& $1.553$& $1.000$\\
&$\mu^+\mu^-$   & $3.635\times 10^{-4}$ & $5.494\times 10^{-3}$ & $3.621\times 10^{-4}$ & $5.494\times 10^{-3}$& $1.000$\\
&$s\bar{s}$   & $3.172\times 10^{-4}$ & $4.795\times 10^{-3}$ & $3.208\times 10^{-4}$& $4.867\times 10^{-3}$& $0.985$\\
&$c\bar{c}$   & $9.829\times 10^{-9}$ & $1.486\times 10^{-7}$ & $9.590\times 10^{-9}$ & $1.455\times 10^{-7}$ & $1.021$\\
&$gg$   & $4.477\times 10^{-4}$ & $6.676\times 10^{-3}$ & $4.604\times 10^{-4}$ & $6.985\times 10^{-3}$ & $0.969$\\
&$\gamma\gamma$   & $1.044\times 10^{-7}$ & $1.578\times 10^{-6}$ & $1.075\times 10^{-7}$ & $1.631\times 10^{-6}$ & $0.968$\\
&$Z\gamma$   & $1.034\times 10^{-7}$ & $1.562\times 10^{-6}$ & $1.027\times 10^{-7}$ & $1.558\times 10^{-6}$ & $1.003$\\
&$W^+W^-$& $7.474\times 10^{-4}$ & $1.130\times 10^{-2}$ & $7.447\times 10^{-4}$ & $1.130\times 10^{-2}$ & $1.000$\\
&$ZZ$   & $3.317\times 10^{-4}$ & $5.014\times 10^{-3}$ & $3.305\times 10^{-4}$ & $5.014\times 10^{-3}$ & $1.000$\\
&$hh$   & $3.493\times 10^{-4}$ & $5.280\times 10^{-3}$ & $3.481\times 10^{-7}$ & $5.281\times 10^{-3}$ & $1.000$\\
&$ZA$   & $7.891\times 10^{-7}$ & $1.193\times 10^{-5}$ & $7.869\times 10^{-7}$ & $1.194\times 10^{-5}$ & $0.999$\\
\hline
\multicolumn{2}{c|}{Total width} &  & $\mathbf{15.12}$ & & $\mathbf{15.17}$  & $\mathbf{0.996}$  \\
\hline\hline
$A\to$ & $b\bar{b}$ & $0.8975$ & $12.42$ & $0.8979$& $12.47$& $0.996$\\
&$\tau^+\tau^-$     & $0.1010$ & $1.399$ & $0.1007$& $1.399$& $1.000$\\
&$\mu^+\mu^-$   & $3.573\times 10^{-4}$ & $4.946\times 10^{-3}$ & $3.561\times 10^{-4}$ & $4.946\times 10^{-3}$& $1.000$\\
&$s\bar{s}$   & $3.179\times 10^{-4}$ & $4.401\times 10^{-3}$ & $3.217\times 10^{-4}$& $4.468\times 10^{-3}$& $0.985$\\
&$c\bar{c}$   & $6.514\times 10^{-9}$ & $9.018\times 10^{-8}$ & $6.358\times 10^{-9}$ & $8.830\times 10^{-8}$ & $1.021$\\
&$gg$   & $5.648\times 10^{-4}$ & $7.818\times 10^{-3}$ & $5.741\times 10^{-4}$ & $7.973\times 10^{-3}$ & $0.981$\\
&$\gamma\gamma$   & $1.126\times 10^{-7}$ & $1.559\times 10^{-6}$ & $1.089\times 10^{-7}$ & $1.512\times 10^{-6}$ & $1.031$\\
&$Z\gamma$   & $5.261\times 10^{-8}$ & $7.283\times 10^{-7}$ & $5.238\times 10^{-8}$ & $7.275\times 10^{-7}$ & $1.001$\\
&$Zh$   & $1.957\times 10^{-4}$ & $2.710\times 10^{-3}$ & $1.951\times 10^{-4}$ & $2.710\times 10^{-3}$ & $1.000$\\
\hline
\multicolumn{2}{c|}{Total width} &  & $\mathbf{13.84}$ & & $\mathbf{13.89}$  & $\mathbf{0.997}$  \\\hline\hline
\end{tabular}

%% file: tables/table-br-C.tex
\begin{tabular}{rl|cc|cc|c}
\hline
& &  \multicolumn{2}{c|}{\tt 2HDMC} & \multicolumn{2}{c|}{\tt HDECAY}\\
& &  $\mathrm{BR}$ & $\Gamma$ (GeV) & $\mathrm{BR}$ & $\Gamma$ (GeV) & $\Gamma_{\mathrm{2H}}/\Gamma_{\mathrm{HD}}$\\
\hline
$h \to$ & $b\bar{b}$& $0.3536$ & $8.207\times 10^{-4}$ & $0.3551$& $8.272\times 10^{-4}$ & $0.992$\\
&$\tau^+\tau^-$   & $3.395\times 10^{-2}$ & $7.880\times 10^{-5}$ & $3.383\times 10^{-2}$ & $7.881\times 10^{-5}$& $1.000$\\
&$\mu^+\mu^-$   & $1.202\times 10^{-4}$ & $2.790\times 10^{-7}$ & $1.198\times 10^{-4}$ & $2.791\times 10^{-7}$ & $1.000$\\
&$s\bar{s}$   & $1.311\times 10^{-4}$ & $3.042\times 10^{-7}$ & $1.323\times 10^{-4}$ & $3.028\times 10^{-7}$ & $0.987$\\
&$c\bar{c}$   & $5.211\times 10^{-2}$ & $1.210\times 10^{-4}$ & $5.082\times 10^{-2}$ & $1.184\times 10^{-4}$ & $1.022$\\
&$gg$  & $0.1453$ & $3.374\times 10^{-4}$ & $0.1467$ & $3.417\times 10^{-4}$ & $0.987$\\
&$\gamma\gamma$   & $3.791\times 10^{-3}$ & $8.799\times 10^{-6}$ & $3.761\times 10^{-3}$ & $8.761\times 10^{-6}$ & $1.004$\\
&$Z\gamma$   & $2.627\times 10^{-3}$ & $6.098\times 10^{-6}$ & $2.681\times 10^{-3}$ & $6.099\times 10^{-6}$ & $1.000$\\
&$W^+W^-$& $0.3630$ & $8.427\times 10^{-4}$ & $0.3617$ & $8.426\times 10^{-4}$ & $1.000$\\
&$ZZ$   & $4.535\times 10^{-2}$ & $1.053\times 10^{-4}$ & $4.519\times 10^{-2}$ & $1.053\times 10^{-4}$ & $1.000$\\
\hline
\multicolumn{2}{c|}{Total width}& & $\bf{2.321\times 10^{-3}}$ & & $\mathbf{2.329\times 10^{-3}}$ & $\mathbf{0.996}$  \\
\hline\hline
$H\to$ & $b\bar{b}$ & $0.7671$ & $0.6903$ & $0.7675$& $0.6926$& $0.997$\\
&$\tau^+\tau^-$     & $9.273\times 10^{-2}$& $8.345\times 10^{-2}$& $9.247\times 10^{-2}$& $8.344\times 10^{-2}$& $1.000$\\
&$\mu^+\mu^-$   & $3.279\times 10^{-4}$ & $2.951\times 10^{-4}$ & $3.270\times 10^{-4}$ & $2.951\times 10^{-4}$& $1.000$\\
&$s\bar{s}$   & $2.717\times 10^{-4}$ & $2.445\times 10^{-4}$ & $2.751\times 10^{-4}$& $2.483\times 10^{-4}$& $0.985$\\
&$c\bar{c}$   & $1.044\times 10^{-6}$ & $9.392\times 10^{-7}$ & $1.019\times 10^{-6}$ & $9.195\times 10^{-7}$ & $1.021$\\
&$t\bar{t}$   & $1.416\times 10^{-2}$ & $1.274\times 10^{-2}$ & $1.436\times 10^{-2}$ & $1.296\times 10^{-2}$ & $0.983$\\
&$gg$   & $5.203\times 10^{-4}$ & $4.682\times 10^{-4}$ & $5.358\times 10^{-4}$ & $4.835\times 10^{-4}$ & $0.968$\\
&$\gamma\gamma$   & $1.188\times 10^{-6}$ & $1.069\times 10^{-6}$ & $1.319\times 10^{-6}$ & $1.190\times 10^{-6}$ & $0.898$\\
&$Z\gamma$   & $1.764\times 10^{-6}$ & $1.588\times 10^{-6}$ & $1.759\times 10^{-6}$ & $1.587\times 10^{-6}$ & $1.000$\\
&$W^+W^-$& $3.586\times 10^{-2}$ & $3.227\times 10^{-2}$ & $3.576\times 10^{-2}$ & $3.227\times 10^{-2}$ & $1.000$\\
&$ZZ$   & $1.672\times 10^{-2}$ & $1.504\times 10^{-2}$ & $1.667\times 10^{-2}$ & $1.504\times 10^{-2}$ & $1.000$\\
&$hh$   & $7.229\times 10^{-2}$ & $6.505\times 10^{-2}$ & $7.209\times 10^{-2}$ & $6.505\times 10^{-2}$ & $1.000$\\
\hline
\multicolumn{2}{c|}{Total width} &  & $\mathbf{0.8999}$ & & $\mathbf{0.9024}$  & $\mathbf{0.997}$  \\
\hline\hline
$A\to$ & $b\bar{b}$ & $0.6117$ & $0.8253$ & $0.6138$& $0.8274$& $0.997$\\
&$\tau^+\tau^-$     & $7.679\times 10^{-2}$& $0.1036$ & $7.684\times 10^{-2}$& $0.1036$& $1.000$\\
&$\mu^+\mu^-$   & $2.716\times 10^{-4}$ & $3.663\times 10^{-4}$ & $2.717\times 10^{-4}$ & $3.663\times 10^{-3}$& $1.000$\\
&$s\bar{s}$   & $2.166\times 10^{-4}$ & $2.923\times 10^{-4}$ & $2.200\times 10^{-4}$& $2.966\times 10^{-4}$& $0.986$\\
&$c\bar{c}$   & $2.774\times 10^{-6}$ & $3.742\times 10^{-6}$ & $2.717\times 10^{-6}$ & $3.663\times 10^{-6}$ & $1.022$\\
&$t\bar{t}$ & $0.1690$ & $0.2279$ & $0.1668$ & $0.2248$& $1.014$\\
&$gg$   & $8.384\times 10^{-4}$ & $1.131\times 10^{-3}$ & $8.560\times 10^{-4}$ & $1.154\times 10^{-3}$ & $0.980$\\
&$\gamma\gamma$   & $1.761\times 10^{-6}$ & $2.376\times 10^{-6}$ & $1.860\times 10^{-6}$ & $2.507\times 10^{-6}$ & $0.948$\\
&$Z\gamma$   & $4.851\times 10^{-7}$ & $6.545\times 10^{-7}$ & $4.854\times 10^{-7}$ & $6.543\times 10^{-7}$ & $1.000$\\
&$Zh$   & $4.426\times 10^{-2}$ & $5.971\times 10^{-2}$ & $4.428\times 10^{-2}$ & $5.969\times 10^{-2}$ & $1.000$\\
&$ZH$   & $9.694\times 10^{-2}$ & $0.1308$ & $9.699\times 10^{-2}$ & $0.1307$ &   $1.000$\\
\hline
\multicolumn{2}{c|}{Total width} &  & $\mathbf{1.349}$ & & $\mathbf{1.348}$  & $\mathbf{1.001}$  \\\hline\hline
\end{tabular}

%% file: 2hdmreco.bbl
\begin{thebibliography}{10}
%
%

\bibitem{Branco:2011iw} 
  G.~C.~Branco, P.~M.~Ferreira, L.~Lavoura, M.~N.~Rebelo, M.~Sher and J.~P.~Silva,
  Phys.\ Rept.\  {\bf 516}, 1 (2012)
  [arXiv:1106.0034 [hep-ph]].

\bibitem{Harlander:2002wh}
  R.~V.~Harlander and W.~B.~Kilgore,
  Phys.\ Rev.\ Lett.\  {\bf 88} (2002) 201801
  [hep-ph/0201206].

\bibitem{Anastasiou:2002yz}
  C.~Anastasiou and K.~Melnikov,
  Nucl.\ Phys.\ B {\bf 646} (2002) 220
  [hep-ph/0207004].

\bibitem{Ravindran:2003um}
  V.~Ravindran, J.~Smith and W.~L.~van Neerven,
  Nucl.\ Phys.\ B {\bf 665} (2003) 325
  [hep-ph/0302135].

\bibitem{Graudenz:1992pv}
  D.~Graudenz, M.~Spira and P.~M.~Zerwas,
  Phys.\ Rev.\ Lett.\  {\bf 70} (1993) 1372.

\bibitem{Spira:1993bb}
  M.~Spira, A.~Djouadi, D.~Graudenz and P.~M.~Zerwas,
  Phys.\ Lett.\ B {\bf 318} (1993) 347.

\bibitem{Spira:1995rr}
  M.~Spira, A.~Djouadi, D.~Graudenz and P.~M.~Zerwas,
  Nucl.\ Phys.\ B {\bf 453} (1995) 17
  [hep-ph/9504378].

\bibitem{Marzani:2008ih}
  S.~Marzani, R.~D.~Ball, V.~Del Duca, S.~Forte and A.~Vicini,
  Nucl.\ Phys.\ Proc.\ Suppl.\  {\bf 186} (2009) 98
  [arXiv:0809.4934 [hep-ph]].

\bibitem{Harlander:2009my}
  R.~V.~Harlander, H.~Mantler, S.~Marzani and K.~J.~Ozeren,
  Eur.\ Phys.\ J.\ C {\bf 66} (2010) 359
  [arXiv:0912.2104 [hep-ph]].

\bibitem{Harlander:2009mq}
  R.~V.~Harlander and K.~J.~Ozeren,
  JHEP {\bf 0911} (2009) 088
  [arXiv:0909.3420 [hep-ph]].

\bibitem{Pak:2009dg}
  A.~Pak, M.~Rogal and M.~Steinhauser,
  JHEP {\bf 1002} (2010) 025
  [arXiv:0911.4662 [hep-ph]].

\bibitem{Pak:2011hs}
  A.~Pak, M.~Rogal and M.~Steinhauser,
  JHEP {\bf 1109} (2011) 088
  [arXiv:1107.3391 [hep-ph]].

\bibitem{Harlander:2012pb}
  R.~V.~Harlander, S.~Liebler and H.~Mantler,
  Comp.\ Phys.\ Commun.\ {\bf 184} (2013) 1605
  [arXiv:1212.3249 [hep-ph]].

\bibitem{higlu}
  M.~Spira,
  hep-ph/9510347
and
  Nucl.\ Instrum.\ Meth.\ A {\bf 389} (1997) 357.

\bibitem{Catani:2003zt}
  S.~Catani, D.~de Florian, M.~Grazzini and P.~Nason,
  JHEP {\bf 0307} (2003) 028
  [hep-ph/0306211].

\bibitem{Ravindran}
  V.~Ravindran,
  Nucl.\ Phys.\ B {\bf 746} (2006) 58
and
  Nucl.\ Phys.\ B {\bf 752} (2006) 173.

\bibitem{Anastasiou:2008tj}
  C.~Anastasiou, R.~Boughezal and F.~Petriello,
  JHEP {\bf 0904} (2009) 003
  [arXiv:0811.3458 [hep-ph]].

\bibitem{Moch:2005ky}
  S.~Moch and A.~Vogt,
  Phys.\ Lett.\ B {\bf 631} (2005) 48
  [hep-ph/0508265].

\bibitem{Ball:2013bra}
  R.~D.~Ball, M.~Bonvini, S.~Forte, S.~Marzani and G.~Ridolfi,
  Nucl.\ Phys.\ B {\bf 874} (2013) 746
  [arXiv:1303.3590 [hep-ph]].

\bibitem{Djouadi:1994ge}
  A.~Djouadi and P.~Gambino,
  Phys.\ Rev.\ Lett.\  {\bf 73} (1994) 2528
  [hep-ph/9406432].

\bibitem{Actis:2008ug}
  S.~Actis, G.~Passarino, C.~Sturm and S.~Uccirati,
  Phys.\ Lett.\ B {\bf 670} (2008) 12
  [arXiv:0809.1301 [hep-ph]].

\bibitem{Degrassi:2004mx}
  G.~Degrassi and F.~Maltoni,
  Phys.\ Lett.\ B {\bf 600} (2004) 255
  [hep-ph/0407249].

\bibitem{Aglietti:2004nj}
  U.~Aglietti, R.~Bonciani, G.~Degrassi and A.~Vicini,
  Phys.\ Lett.\ B {\bf 595} (2004) 432
  [hep-ph/0404071].

\bibitem{Dicus:1988cx}
  D.~A.~Dicus and S.~Willenbrock,
  Phys.\ Rev.\ D {\bf 39} (1989) 751.

\bibitem{Campbell:2002zm}
  J.~M.~Campbell, R.~K.~Ellis, F.~Maltoni and S.~Willenbrock,
  Phys.\ Rev.\ D {\bf 67} (2003) 095002
  [hep-ph/0204093].

\bibitem{Dicus:1998hs}
  D.~Dicus, T.~Stelzer, Z.~Sullivan and S.~Willenbrock,
  Phys.\ Rev.\ D {\bf 59} (1999) 094016
  [hep-ph/9811492].

\bibitem{Maltoni:2003pn}
  F.~Maltoni, Z.~Sullivan and S.~Willenbrock,
  Phys.\ Rev.\ D {\bf 67} (2003) 093005
  [hep-ph/0301033].

\bibitem{Dittmaier:2003ej}
  S.~Dittmaier, M.~Kr\"amer and M.~Spira,
  Phys.\ Rev.\ D {\bf 70} (2004) 074010
  [hep-ph/0309204].

\bibitem{Dawson:2003kb}
  S.~Dawson, C.~B.~Jackson, L.~Reina and D.~Wackeroth,
  Phys.\ Rev.\ D {\bf 69} (2004) 074027
  [hep-ph/0311067].

\bibitem{Harlander:2003ai}
  R.~V.~Harlander and W.~B.~Kilgore,
  Phys.\ Rev.\ D {\bf 68} (2003) 013001
  [hep-ph/0304035].

\bibitem{4fsgrids}
The files {\tt 4f\_pseudoscalar\_7.root}, {\tt 4f\_pseudoscalar\_8.root},
{\tt 4f\_scalar\_7.root} and {\tt 4f\_scalar\_8.root} containing the 4FS MSSM numbers can be found at \\
{\tt https://svnweb.cern.ch/cern/wsvn/lhchiggsxs/repository/MSSM/Neutral
Higgs/machinery\_with\_SusHI/root\_files/?\#a6f5d38d4636ed77fc65dc5f03a8bdcb5};
\\
{\tt https://twiki.cern.ch/twiki/bin/view/LHCPhysics/MSSMNeutral}.

\bibitem{santander}
  R.~Harlander, M.~Kr\"amer and M.~Schumacher,
  arXiv:1112.3478 [hep-ph].

\bibitem{ptfix}
  R.~K.~Ellis, I.~Hinchliffe, M.~Soldate and J.~J.~van der Bij,
  Nucl.\ Phys.\ B {\bf 297} (1988) 221;
  U.~Baur and E.W.N.~Glover,
  Nucl.\ Phys.\ B {\bf 339} (1990) 38;
  O.~Brein and W.~Hollik,
  Phys.\ Rev.\ D {\bf 68} (2003) 095006;
  U.~Langenegger, M.~Spira, A.~Starodumov and P.~Tr\"ub,
  JHEP {\bf 0606} (2006) 035.

\bibitem{ptnlo}
  C.~R.~Schmidt,
  Phys.\ Lett.\ B {\bf 413} (1997) 391;
  D.~de Florian, M.~Grazzini and Z.~Kunszt,
  Phys.\ Rev.\ Lett.\  {\bf 82} (1999) 5209;
  V.~Ravindran, J.~Smith and W.~L.~Van Neerven,
  Nucl.\ Phys.\ B {\bf 634} (2002) 247;
  C.~J.~Glosser and C.~R.~Schmidt,
  JHEP {\bf 0212} (2002) 016;
  C.~Anastasiou, K.~Melnikov and F.~Petriello,
  Phys.\ Rev.\ Lett.\  {\bf 93} (2004) 262002
and
  Nucl.\ Phys.\ B {\bf 724} (2005) 197.

\bibitem{ptnnlo}
  R.~Boughezal, F.~Caola, K.~Melnikov, F.~Petriello and M.~Schulze,
  JHEP {\bf 1306} (2013) 072.

\bibitem{ptnlomt}
  R.V.~Harlander, T.~Neumann, K.J.~Ozeren and M.~Wiesemann,
  JHEP {\bf 1208} (2012) 139.

\bibitem{ptresum}
  S.~Catani, E.~D'Emilio and L.~Trentadue,
  Phys.\ Lett.\ B {\bf 211} (1988) 335;
  I.~Hinchliffe and S.F.~Novaes,
  Phys.\ Rev.\ D {\bf 38} (1988) 3475;
  R.~P.~Kauffman,
  Phys.\ Rev.\ D {\bf 44} (1991) 1415
and
  Phys.\ Rev.\ D {\bf 45} (1992) 1512;
  C.~Balazs and C.P.~Yuan,
  Phys.\ Lett.\ B {\bf 478} (2000) 192;
  E.~L.~Berger and J.W.~Qiu,
  Phys.\ Rev.\ D {\bf 67} (2003) 034026;
  A.~Kulesza and W.~J.~Stirling,
  JHEP {\bf 0312} (2003) 056;
  A.~Kulesza, G.~Sterman and W.~Vogelsang,
  Phys.\ Rev.\ D {\bf 69} (2004) 014012;
  A.~Gawron and J.~Kwiecinski,
  Phys.\ Rev.\ D {\bf 70} (2004) 014003;
  G.~Watt, A.~D.~Martin and M.~G.~Ryskin,
  Phys.\ Rev.\ D {\bf 70} (2004) 014012
  [Erratum-ibid.\ D {\bf 70} (2004) 079902];
  A.~V.~Lipatov and N.~P.~Zotov,
  Eur.\ Phys.\ J.\ C {\bf 44} (2005) 559;
  D.~de Florian and M.~Grazzini,
  Phys.\ Rev.\ Lett.\  {\bf 85} (2000) 4678;
  Nucl.\ Phys.\ B {\bf 616} (2001) 247;
  S.~Catani, D.~de Florian and M.~Grazzini,
  Nucl.\ Phys.\ B {\bf 596} (2001) 299;
  G.~Bozzi, S.~Catani, D.~de Florian and M.~Grazzini,
  Phys.\ Lett.\ B {\bf 564} (2003) 65;
  G.~Bozzi, S.~Catani, D.~de Florian and M.~Grazzini,
  Nucl.\ Phys.\ B {\bf 737} (2006) 73
and
  Nucl.\ Phys.\ B {\bf 791} (2008) 1;
  D.~de Florian, G.~Ferrera, M.~Grazzini and D.~Tommasini,
  JHEP {\bf 1111} (2011) 064.

\bibitem{Alwall:2011cy}
  J.~Alwall, Q.~Li and F.~Maltoni,
  Phys.\ Rev.\ D {\bf 85} (2012) 014031
  [arXiv:1110.1728 [hep-ph]].

\bibitem{Bagnaschi:2011tu}
  E.~Bagnaschi, G.~Degrassi, P.~Slavich and A.~Vicini,
  JHEP {\bf 1202} (2012) 088
  [arXiv:1111.2854 [hep-ph]].
%

\bibitem{Mantler:2012bj}
  H.~Mantler and M.~Wiesemann,
  Eur.\ Phys.\ J.\ C {\bf 73} (2013) 2467
  [arXiv:1210.8263 [hep-ph]].

\bibitem{grazzinicodes} \url{http://theory.fi.infn.it/grazzini/codes.html}

\bibitem{Grazzini:2013mca}
  M.~Grazzini and H.~Sargsyan,
  JHEP {\bf 1309} (2013) 129
  [arXiv:1306.4581 [hep-ph]].

\bibitem{Banfi:2013eda}
  A.~Banfi, P.~F.~Monni and G.~Zanderighi,
  arXiv:1308.4634 [hep-ph].

\bibitem{Frixione:2002ik}
  S.~Frixione and B.~R.~Webber,
  JHEP {\bf 0206} (2002) 029
  [hep-ph/0204244].

\bibitem{mcatnlo}
{\tt http://www.hep.phy.cam.ac.uk/theory/webber/MCatNLO}

\bibitem{bbhlh}
  J.~M.~Campbell, S.~Dawson, S.~Dittmaier, C.~Jackson, M.~Kr\"amer,
  F.~Maltoni, L.~Reina, M.~Spira, D.~Wackeroth and S.~Willenbrock,
  hep-ph/0405302.

\bibitem{Harlander:2010cz}
  R.~V.~Harlander, K.~J.~Ozeren and M.~Wiesemann,
  Phys.\ Lett.\ B {\bf 693} (2010) 269
  [arXiv:1007.5411 [hep-ph]].

\bibitem{mcfm}
\url{http://mcfm.fnal.gov/}

\bibitem{Harlander:2011fx}
  R.~Harlander and M.~Wiesemann,
  JHEP {\bf 1204} (2012) 066
  [arXiv:1111.2182 [hep-ph]].

\bibitem{Buehler:2012cu}
  S.~B\"uhler, F.~Herzog, A.~Lazopoulos and R.~M\"uller,
  JHEP {\bf 1207} (2012) 115
  [arXiv:1204.4415 [hep-ph]].

\bibitem{Belyaev:2005bs}
  A.~Belyaev, P.~M.~Nadolsky and C.~-P.~Yuan,
  JHEP {\bf 0604} (2006) 004
  [hep-ph/0509100].

\bibitem{amcnlo} \url{http://amcatnlo.web.cern.ch/amcatnlo/}

\bibitem{Eriksson:2009ws} 
  D.~Eriksson, J.~Rathsman and O.~St{\aa}l,
  Comput.~Phys.~Commun.~{\bf 181}, 189 (2010)
  [arXiv:0902.0851 [hep-ph]];
  D.~Eriksson, J.~Rathsman and O.~St{\aa}l,
  Comput.\ Phys.\ Commun.\  {\bf 181}, 833 (2010).

  \bibitem{Djouadi:1997yw} 
  A.~Djouadi, J.~Kalinowski and M.~Spira,
  Comput.\ Phys.\ Commun.\  {\bf 108}, 56 (1998)
  [hep-ph/9704448];
  A.~Djouadi, M.~M.~M\"uhlleitner and M.~Spira,
  Acta Phys.\ Polon.\ B {\bf 38} (2007) 635
  [hep-ph/0609292].

\bibitem{fortsch}
M.~Spira,
  Fortsch.\ Phys.\  {\bf 46} (1998) 203
  [hep-ph/9705337];
A.~Djouadi,
  Phys.\ Rept.\  {\bf 459} (2008) 1
  [hep-ph/0503173].

\bibitem{Davidson:2005cw} 
  S.~Davidson and H.~E.~Haber,
  Phys.\ Rev.\ D {\bf 72}, 035004 (2005)
  [Erratum-ibid.\ D {\bf 72}, 099902 (2005)]
  [hep-ph/0504050].

\bibitem{Barger:1989fj} 
  V.~D.~Barger, J.~L.~Hewett and R.~J.~N.~Phillips,
  Phys.\ Rev.\ D {\bf 41}, 3421 (1990).

\bibitem{Bechtle:2013wla} 
  P.~Bechtle, O.~Brein, S.~Heinemeyer, G.~Weiglein and K.~E.~Williams,
  Comput.\ Phys.\ Commun.\  {\bf 181}, 138 (2010)
  [arXiv:0811.4169 [hep-ph]];
  P.~Bechtle, O.~Brein, S.~Heinemeyer, G.~Weiglein and K.~E.~Williams,
  Comput.\ Phys.\ Commun.\  {\bf 182}, 2605 (2011)
  [arXiv:1102.1898 [hep-ph]];
  P.~Bechtle, O.~Brein, S.~Heinemeyer, O.~St{\aa}l, T.~Stefaniak, G.~Weiglein and K.~E.~Williams,
  arXiv:1311.0055 [hep-ph].

\bibitem{nearthreshold}
E. Braaten and J. P. Leveille, Phys. Rev. D {\bf 22} (1980) 715; 
N. Sakai, Phys. Rev. D {\bf 22} (1980) 2220; 
T. Inami and T. Kubota, Nucl. Phys. B {\bf 179} (1981) 171; 
M. Drees and K.--I. Hikasa, Phys. Rev. D {\bf 41} (1990) 1547; 
M. Drees and K.--I. Hikasa, Phys. Lett. B {\bf 240} (1990) 455
[Erratum-ibid. B {\bf 262} (1991) 497].

\bibitem{abovethreshold}
S. G. Gorishnii, A. L. Kataev, S. A. Larin and L. R. Surguladze, Mod. Phys. Lett. A
{\bf 5} (1990) 2703; 
S. G. Gorishnii, A. L. Kataev, S. A. Larin and L. R. Surguladze,
Phys.Rev. D {\bf 43} (1991) 1633; 
A. L. Kataev and V. T. Kim, Mod. Phys. Lett. A {\bf 9} (1994) 1309;
S. G. Gorishnii, A. L. Kataev and S. A. Larin,
Sov. J. Nucl. Phys. {\bf 40} (1984) 329 [Yad. Fiz. 40 (1984) 517]; 
L. R. Surguladze, Phys. Lett. B {\bf 341} (1994) 60 [hep-ph/9405325];
S. A. Larin, T. van Ritbergen and J. A. M. Vermaseren, Phys. Lett. B
{\bf 362} (1995) 134 [hep-ph/9506465]; 
K. G. Chetyrkin and A. Kwiatkowski, Nucl. Phys. B {\bf 461} (1996) 3
[hep-ph/9505358]. 

\bibitem{chetyrkin}
K. G. Chetyrkin, Phys. Lett. B {\bf 390} (1997) 309 [hep-ph/9608318].

\bibitem{baikov}
P. A. Baikov, K. G. Chetyrkin and J. H. K\"uhn, Phys. Rev. Lett. {\bf
  96} (2006) 012003 [hep-ph/0511063].

\bibitem{Djouadi:1994gf}
A.~Mendez and A.~Pomarol,
  Phys.\ Lett.\ B {\bf 252} (1990) 461;
C.--S.~Li and R.~J.~Oakes,
  Phys.\ Rev.\ D {\bf 43} (1991) 855;
  A.~Djouadi and P.~Gambino,
  Phys.\ Rev.\ D {\bf 51} (1995) 218
   [Erratum-ibid.\ D {\bf 53} (1996) 4111]
  [hep-ph/9406431].

\bibitem{offshellphi}
A. Djouadi, J. Kalinowski and P.~M. Zerwas, Z. Phys. C {\bf 70} (1996) 437;
S. Moretti and W.~J. Stirling, Phys. Lett. B {\bf 347} (1995) 291 and
(E) B {\bf 366} (1996) 451.

\bibitem{nloggqcd}
  T.~Inami, T.~Kubota and Y.~Okada,
  Z.\ Phys.\ C {\bf 18} (1983) 69;
  A.~Djouadi, M.~Spira and P.~M.~Zerwas,
  Phys.\ Lett.\ B {\bf 264} (1991) 440;
  K.~G.~Chetyrkin, B.~A.~Kniehl and M.~Steinhauser,
  Phys.\ Rev.\ Lett.\  {\bf 79} (1997) 353
  [hep-ph/9705240];
  K.~G.~Chetyrkin, B.~A.~Kniehl, M.~Steinhauser and W.~A.~Bardeen,
  Nucl.\ Phys.\ B {\bf 535}, 3 (1998)
  [hep-ph/9807241].

\bibitem{Chetyrkin:1997un}
  K.~G.~Chetyrkin, B.~A.~Kniehl and M.~Steinhauser,
  Nucl.\ Phys.\ B {\bf 510} (1998) 61
  [hep-ph/9708255].

\bibitem{Baikov:2006ch}
  P.~A.~Baikov and K.~G.~Chetyrkin,
  Phys.\ Rev.\ Lett.\  {\bf 97} (2006) 061803
  [hep-ph/0604194].

\bibitem{Djouadi:1993ji}
A. Djouadi, M. Spira and P.~M. Zerwas, Phys. Lett. B {\bf 311} (1993) 255;

\bibitem{hgagalim}
  H.--Q.~Zheng and D.--D.~Wu,
  Phys.\ Rev.\ D {\bf 42} (1990) 3760;
  A.~Djouadi, M.~Spira, J.~J.~van der Bij and P.~M.~Zerwas,
  Phys.\ Lett.\ B {\bf 257} (1991) 187;
  S.~Dawson and R.~P.~Kauffman,
  Phys.\ Rev.\ D {\bf 47} (1993) 1264;
  A.~Djouadi, M.~Spira and P.~M.~Zerwas,
  Phys.\ Lett.\ B {\bf 311} (1993) 255
  [hep-ph/9305335];
  K.~Melnikov and O.~I.~Yakovlev,
  Phys.\ Lett.\ B {\bf 312} (1993) 179
  [hep-ph/9302281]; 
  M.~Inoue, R.~Najima, T.~Oka and J.~Saito,
  Mod.\ Phys.\ Lett.\ A {\bf 9} (1994) 1189;
 J.~Fleischer, O.~V.~Tarasov and V.~O.~Tarasov,
 Phys.\ Lett.\ B {\bf 584} (2004) 294
 [hep-ph/0401090].

\bibitem{hgagaqcd}
A. Djouadi, M. Spira, J.~J. van der Bij and P.~M. Zerwas, Phys.~Lett.~B {\bf
  257} (1991) 187;
K. Melnikov and O. Yakovlev, Phys.~Lett.~B {\bf 312} (1993) 179;
M. Inoue, R. Najima, T. Oka and J. Saito, Mod. Phys. Lett.~A {\bf 9} (1994)
1189.

\bibitem{nloZga}
  M.~Spira, A.~Djouadi and P.~M.~Zerwas,
  Phys.\ Lett.\ B {\bf 276} (1992) 350.

\bibitem{Cahn:1988ru}
  R.N.~Cahn,
  Rept.\ Prog.\ Phys.\  {\bf 52} (1989) 389.

\bibitem{Dittmaier:2011ti}
  S.~Dittmaier {\it et al.}  [LHC Higgs Cross Section Working Group Collaboration],
  arXiv:1101.0593 [hep-ph].

\bibitem{pdf4lhc}
PDF4LHC steering committee, {\tt http://www.hep.ucl.ac.uk/pdf4lhc/}.

\end{thebibliography}
